\documentclass[12pt,twoside]{article}
\usepackage{amsmath,mathrsfs,bm,amssymb,color}

\usepackage[dvips]{graphicx}
\usepackage{float}
\usepackage{latexsym}
\catcode`\@=11
\def\newsymbol#1#2#3#4#5{\let\next@\relax%
 \ifnum#2=\@ne\else%
 \ifnum#2=\tw@\let\next@\msyfam@\fi\fi%
 \mathchardef#1="#3\next@#4#5}
\def\mathhexbox@#1#2#3{\relax%
 \ifmmode\mathpalette{} {\m@th\mnnathchar"#1#2#3}
 \else\leavevmode\hbox{$\m@th\mathchar"#1#2#3$}\fi}
\font\tenmsy=msbm10
\font\sevenmsy=msbm7
\font\fivemsy=msbm5
\newfam\msyfam
\textfont\msyfam=\tenmsy
\scriptfont\msyfam=\sevenmsy
\scriptscriptfont\msyfam=\fivemsy
\edef\msyfam@{\hexnumber@\msyfam}
\def\mathbb #1{\fam\msyfam\relax#1}
\catcode`\@=\active

\newtheorem{theorem}{Theorem}[section]
\newtheorem{proposition}[theorem]{Proposition}
\newtheorem{lemma}[theorem]{Lemma}
\newtheorem{corollary}[theorem]{Corollary}
\newtheorem{definition}[theorem]{Definition}

\newtheorem{remark}[theorem]{Remark}

\newcommand{\EE}{{\mathbb E}_P}
\newcommand{\EEE}{{\mathbb  E}_{\mu_{\rm E}}}

\newcommand{\KI}{K(f)}

\newcommand{\ee}[1]{\Ebb_{\ggg} \!\!\lkkk #1 \rkkk}
\newcommand{\EEEE}{\Ebb_{\ggg}\!}

\newcommand{\Ebb}{{\mathbb{E}}}

\newcommand{\bd}[1]{\begin{definition}\label{#1}}
\newcommand{\ed}{\end{definition}}
\newcommand{\La}{\Lambda}

\renewcommand{\d}{\displaystyle}
\newcommand{\eps}{\varepsilon}

\newcommand{\bl}[1]{\begin{lemma}\label{#1}}
\newcommand{\el}{\end{lemma}}
\newcommand{\bc}[1]{\begin{corollary}\label{#1}}
\newcommand{\ec}{\end{corollary}}
\newcommand{\bt}[1]{\begin{theorem}\label{#1}}
\newcommand{\et}{\end{theorem}}
\newcommand{\bp}[1]{\begin{proposition}\label{#1}}
\newcommand{\ep}{\end{proposition}}
\newcommand{\br}[1]{\begin{remark}\label{#1}}
\newcommand{\er}{\end{remark}}
\newcommand{\zz}{{\mathbb  Z}_2}
\renewcommand{\P}{{\cal W}}

\newcommand{\eq}[1]{\begin{equation}\label{#1}}
\newcommand{\en}{\end{equation}}
\newcommand{\eqn}{\begin{eqnarray*}}
\newcommand{\enn}{\end{eqnarray*}}
\newcommand{\eqnn}{\begin{eqnarray}}
\newcommand{\ennn}{\end{eqnarray}}
\newcommand{\proof}{{\noindent \it Proof:\ }}
\newcommand{\qed}{\hfill $\Box$\par\medskip}
\newcommand{\BR}{{{\mathbb  R}^d }}
\newcommand{\bi}{\begin{description}}
\newcommand{\ei}{\end{description} }
\newcommand{\CC}{{{\mathbb  C}}}
\newcommand{\RR}{{\mathbb  R}}

\providecommand{\seq}[1]{(#1_n)_{n\in \mathbb {N}}}
\newcommand{\pro}[1]{(#1_t)_{t\in\RR}}

\newcommand{\limn}{\lim_{n\rightarrow\infty}}

\newcommand{\wick}[1]{{:\!\! #1 \!\!:}}
\newcommand{\kak}[1]{(\ref{#1})}

\renewcommand{\ggg}{\mu_{\rm g}}

\newcommand{\LR}{{L^2(\BR)}}
\newcommand{\LZ}{{L^2(\zz)}}

\newcommand{\ob}{\Omega_{\rm b}}
\newcommand{\LRT}{{L^2(\RR^{d +1})}}

\newcommand{\ms}{\mathscr }
\newcommand{\nnn}{{\ms N}}
\newcommand{\FFF}{{\ms Y}}

\newcommand{\fff}{{\ms F}}
\newcommand{\is}{\inf{\spec}}

\newcommand{\ov}[1]{\overline{#1}}

\newcommand{\lk}{\left(}
\newcommand{\rk}{\right)}
\newcommand{\lkk}{\left\{}
\newcommand{\rkk}{\right\}}
\newcommand{\lkkk}{\left[}
\newcommand{\rkkk}{\right]}

\newcommand{\add}{a^{\dagger}}
\newcommand{\ass}{a^\sharp}

\newcommand{\PF}{{H}}
\newcommand{\SB}{H_{\rm SB}}

\newcommand{\hf}{H_{\rm f}}

\newcommand{\tPF}{{\tilde H}}

\newcommand{\grvh}{\varphi_{\rm vH}}
\newcommand{\vh}{H_{\rm vH}}
\newcommand{\phib}{\phi_{\rm b}}

\newcommand{\gr}{\varphi_{\rm g}}
\newcommand{\grsb}{\varphi_{\rm SB}}

\newcommand{\grt}{\frac{\Phi_T}{\|\Phi_T\|}}

\def\bbbone{{\mathchoice {\rm 1\mskip-4mu l} {\rm 1\mskip-4mu l}
{\rm 1\mskip-4.5mu l} {\rm 1\mskip-5mu l}}}
\def\one{\bbbone}

\newcommand {\onee}{\one_\hhh}

\newcommand {\n}  {N}
\newcommand {\nn}  {d\Gamma(\rho(-i\nabla))}

\newcommand{\half}{\frac{1}{2}}
\newcommand{\han}{{1/2}}
\newcommand{\field}{\phi_{\rm b}}

\newcommand{\vvv}[1]{\left[
\!\!\!\begin{array}{c}#1\end{array}\!\!\!\right]}
\newcommand{\mmm}[4]
{\left[ \!\!\!\begin{array}{cc}#1&#2\\
#3&#4\end{array}\!\!\!\right]}

\newcommand{\spec}{{\rm Spec}}

\newcommand{\s}{\sigma}
\newcommand{\hhh}{{\ms H}}

\newcommand{\non}{\nonumber}

\makeatletter
\@addtoreset{equation}{section}
\makeatother

\title
{\sc Spin-Boson Model through a Poisson-Driven Stochastic Process}

\author{
\small Masao Hirokawa \\
{\it \small Department of Mathematics, Okayama University} \\[-0.7ex]
{\it \small Okayama, 700-8530, Japan} \\[-0.7ex]
{\small {\tt hirokawa@math.okayama-u.ac.jp }} \\[0.3cm]
\small Fumio Hiroshima\\
{\small\it Faculty of Mathematics, Kyushu University}    \\[-0.7ex]
{\small\it Fukuoka, 819-0395, Japan}      \\[-0.7ex]
{\small  {\tt hiroshima@math.kyushu-u.ac.jp}}\\[0.3cm]
\small J\'ozsef L\H{o}rinczi \\
{\small\it School of Mathematics, Loughborough University}    \\[-0.7ex]
{\small\it Loughborough LE11 3TU, United Kingdom}      \\[-0.7ex]
{\small  {\tt J.Lorinczi@lboro.ac.uk}  } \\[-0.7ex]}

\bigskip\bigskip\bigskip

\date{}

\begin{document}
\maketitle
\setlength{\baselineskip}{17pt}

\bigskip\bigskip\bigskip

\begin{abstract}
\noindent
We give a functional integral representation of the semigroup generated by the spin-boson Hamiltonian
by making use of a Poisson point process and a Euclidean field. We present a method of constructing
Gibbs path measures indexed by the full real line which can be applied also to more general stochastic
processes with jump discontinuities. Using these tools we then show existence and uniqueness of the
ground state of the spin-boson, and analyze ground state properties. In particular, we prove
super-exponential decay of the number of bosons, Gaussian decay of the field operators, derive
expressions for the positive integer, fractional and exponential moments of the field operator, and
discuss the field fluctuations in the ground state.

\end{abstract}

\newpage
\section{Spin-boson model}
\subsection{Introduction}
\label{s11}
Gibbs measures constructed on the space of continuous paths of a random process proved to play an important
role in studying ground state properties of Hamiltonians in quantum field theory (\cite[p.298-p.324]{lhb11} and
\cite{lm01,lms02, l02, bhlms02,bh09,gl09,ghl12}). Such random processes are obtained by conditioning Brownian
motion with respect to an external and a pair interaction potential. In this setting Gibbs measures are obtained
as weak limits of sequences of Gibbs measures indexed by the bounded intervals of the real line by using
pre-compactness or tightness arguments.

In the present paper we extend this strategy to construct Gibbs measures on paths of a random process with
jump discontinuities (c\`adl\`ag paths) associated with the Hamilton operator of the spin-boson model:
\eq{sb100}
H=-\eps\s_x\otimes\one+\one\otimes\hf+\alpha\s_z\otimes\phib(\hat h)
\en
with a view of studying spectral properties of this Hamiltonian. Here $\eps>0$ and $\alpha\in\RR$ are parameters,
$\sigma_x, \sigma_z$ are Pauli matrices, $\hf$ is the free field Hamiltonian, $\phib(\hat h)$ is the field
operator in Fock space $\fff$, and $\hat h$ is a form factor (see the details below). One of the merits of this
approach is that it allows to carry through this analysis in a non-perturbative way. While in \cite{spo89,ssw90}
the spectral properties of the spin-boson model are discussed through a measure on the space of paths with jump
discontinuities, no attention was paid to constructing Gibbs measures.

As it will be seen below, in the case of the spin-boson model Gibbs measures involve densities dependent on
a pair interaction potential alone, and no external potential contribution. We stress that, in contrast, in
the case of Brownian motion under zero external potential and non-zero pair interaction potential even the
very existence of Gibbs measures is poorly understood. A rigorous study of Gibbs measures with an external
potential but without pair interaction on c\`adl\`ag paths was begun in \cite{kl12a}, considering them with
respect to fractional $P(\phi)_1$-processes, i.e., stable processes under an external potential.

For Hamiltonians with spin or a non-local kinetic term Feynman-Kac-type formulae can be derived by using
L\'evy processes \cite{hir12,hl08,lm12,hil12a,hil12b,kl12b}. There are few rigorous results on quantum
models with spin using functional integration methods. In \cite{hl08} we derived such a formula for the
heat semigroup generated by a quantum field operator with spin by making use of a Euclidean quantum field
and a Poisson process. In \cite{hil12b} we developed similar methods for relativistic Schr\"odinger operators,
allowing to obtain results on the decay of ground states.

In this paper we derive a Feynman-Kac-type formula for the semigroup generated by the spin-boson Hamiltonian.
The spin-boson model is a much studied variant of the Caldeira-Leggett model describing a two-state quantum
system linearly coupled to a scalar quantum field \cite{lcdfgz87}. Work on the spectral properties of the
spin-boson and related models includes \cite{amrz08, ds85, spo89,ssw90,fn88, hs95,ger96, bs98,ah97,hir99, ger00,hir01,hir02}.
In particular, in \cite{ds85,spo89,ssw90,hs95} stochastic methods were used. Existence or absence  of a
spin-boson ground state was investigated in \cite{ah97, ger00}. Below we obtain existence and uniqueness of
the ground state by a different approach. In \cite{hh12} we apply the method developed in this paper to the
so-called Rabi model, which can be regarded as a single-mode spin-boson model. We also refer to the recent
papers \cite{abd12,hh10}.

In order to study the spin-boson in a stochastic representation we describe the spin states by the set $\zz=
\{-1,+1\}$ and derive a Poisson-driven random process with c\`adl\`ag path space $\ms X = D(\RR,\zz)$, indexed
by the real line and taking values in $\zz$. This will describe the spin-process. The spin-boson Hamiltonian
can be defined as a self-adjoint operator on a Hilbert space $L^2(\zz\times Q)$ instead of $\CC^2\otimes\fff$.

On path space $\ms X$ we are then able to construct a Gibbs measure $\ggg $ associated with the unique ground
state $\gr$ of the spin-boson Hamiltonian. Using this probability measure we represent ground state expectations
for interesting choices of operators ${\cal O}$ in the form
\eq{ex2}
(\gr, {\cal  O}\gr) = \int_{\ms X} f_{{\cal O}} d \ggg,
\en
where $f_{{\cal O}}$ is a function on path space $\ms X$ uniquely associated with ${{\cal O}}$. We will consider
the field operator $\phi(f)$ with test function $f$, and the second quantization $\nn$ of the multiplication
operator by a function $\rho$, and derive path integral representations of expressions of the type
\eq{path1}
{\cal O}=\xi(\s)F(\phi(f))  \quad \mbox{and} \quad  {\cal O} =\xi(\s)e^{-\beta
\nn}
\en
with suitable $\xi:\zz\to\CC$, $F:\RR\to\CC$ and $\rho:\BR\to[0,\infty)$. These cases include
$$
{\cal O}= N^m,\quad {\cal O}=e^{\beta N}, \quad {\cal O}=(-1)^N\quad \mbox{and} \quad\s(-1)^N
$$
for all $m \in \Bbb N$ and $\beta\in\CC$ (in particular, $\beta > 0$), where $N=d\Gamma(\one)$ is the boson
number operator, as well as
$$
{\cal O}=e^{(\beta/2) \phi(f)^2},\quad
{\cal O}= e^{i\beta\phi(f)},\quad
{\cal O}=\phi(f)^n,\quad
{\cal O}=|\phi(f)|^{s}\quad \mbox{and} \quad
{\cal O}=\s \phi(f).
$$
Here $\beta\in\RR$, $n\in{\mathbb N}$ and $0<s<2$.

Specifically, we obtain explicit formulae for the positive integer and fractional order moments and exponential
moments of the field operator, and show that the field fluctuations increase on switching on the coupling between
the spin and the boson field. Moreover, we  show that
$$
\gr\in D(e^{(\beta/2)\phi(f)^2})
$$
for $-\infty <\beta<1/\|f\|^2$ (i.e., Gaussian decay of the field operators) with
$$
\lim_{\beta\uparrow 1/\|f\|^2}\|e^{(\beta/2)\phi(f)^2}\gr\|=\infty.
$$
As a consequence, we obtain another representation of the ground state. Recall that when $\eps=0$, the spin boson
Hamiltonian can be diagonalized so that each matrix element is a van Hove Hamiltonian (see also below). Then it is
trivial to see that
$$
(\gr, F(\phi(f))\gr)={(\grvh, F(\phib(\hat f))\grvh)},
$$
where $\grvh$ is the ground state of the van Hove Hamiltonian in $\fff$. Here we show a similar representation for
the case of $\eps\not=0$, i.e., we derive
$$
(\gr, F(\phi(f))\gr)=\int_{\ms X} {(\grvh(\chi), F(\phib(\hat f))\grvh(\chi))} d\ggg,
$$
where $\grvh(\chi)$ is the ground state of a random van Hove model and $\chi$ is a function of the random path.
This suggests implicitly that $\gr=\int_{\ms X}^\oplus \grvh(\chi) d\ggg$.

Next we discuss ground state properties of second quantized operators. In particular,  we  prove that
$$
\gr\in D(e^{\beta N}), \quad \beta>0
$$
(i.e., super-exponential decay of the number of bosons). Also, we obtain explicit formulae for the moments of the
boson number operator in terms of sums involving coefficients given by the Stirling numbers of the second kind.
Furthermore, although we show that $(\gr, \s(-1)^N\gr)=-1$,  we obtain a positive lower bound on the ground state
functional $(\gr, (-1)^N\gr)$. Finally, we obtain the inequality
$$
(\gr, N\gr)\leq \frac{\alpha}{2}(\gr, \phi(\omega({\rm D})^{-1} h)\gr)\leq
\frac{\alpha^2}{2}\|\hat h/\omega\|^2,\quad {\rm D}=-i\nabla,
$$
relating the mean of the field operator with the expected boson number in the ground state.

These applications to ground state properties are derived from
the main results of this paper, which can
be summarized as follows:
\begin{enumerate}
\item[(1)]
existence and uniqueness of the ground state $\gr$ of the spin-boson Hamiltonian $H$ is shown in Theorem
\ref{existence}
\item[(2)]
a probability measure $\ggg$ on c\`adl\`ag path space $\ms X$ associated with $\gr$ is constructed in
Theorems \ref{gibbs} and \ref{gibbs2}
\item[(3)]
it is shown in Theorem \ref{gibbsmeasure} that   $\ggg$ is a Gibbs measure for a pair interaction potential
\item[(4)]
path integral representations of $(\gr, {\cal O}\gr)$ in terms of $\ggg$ are given in Section 4, in particular,
\begin{enumerate}
\item[(i)]
it is shown that
$$\gr\in D(e^{a  N})\cap  D(e^{b \phi(f)^2}),\quad
(a,b)\in
\RR\times  (-\infty, 1/\|f\|^2)$$ in
Theorem \ref{gaussian decay}, Corollary \ref{gaussian nodecay} and Corollary \ref{1}
\item[(ii)]
it is shown  that
$(\gr, {\cal O}\gr)$ can be represented by the ground state $\grvh(\chi)$ of a van Hove
Hamiltonian  as
$$(\gr, {\cal O}\gr)=\int_{\ms X}(\grvh(\chi), {\cal O}\grvh(\chi))d\ggg $$ in Theorem \ref{rep11}.
\end{enumerate}
\end{enumerate}

We note that path integral representations were already used to a great extent for the Nelson model describing the
interaction of a charged particle with a scalar quantum field, see \cite{bhlms02} and \cite[p.294-p.296]{lhb11}. We
also note that although the result $(\gr, e^{\beta N}\gr)<\infty$ has been established in \cite{gro73} by using
operator theory, our construction here is completely different and rather general. Also, our methods can be applied
to further models involving c\`adl\`ag paths, for instance, the Nelson model with a relativistic kinetic term
$\sqrt{-\Delta+m^2}+V$, which will be done elsewhere.

The paper is organized as follows.
The remainder of Section 1 is devoted to constructing the Feynman-Kac formula of the spin-boson
heat semigroup.
In Section 2 we show that the spin-boson Hamiltonian has a unique ground state $\gr$ if an infrared
regularity condition is satisfied.
In Section 3 we define a Gibbs measure on $\ms X$ for bounded time intervals
associated with the density obtained from the Feynman-Kac representation, and show its local weak convergence to a
path measure in the infinite time limit. We view Theorems \ref{gibbs} and \ref{gibbs2} below to be pivotal results
in this paper.
In Section 4 we derive the expressions of the ground state expectations mentioned above.

\subsection{Definition}
We begin by defining the spin-boson Hamiltonian as a self-adjoint operator on a Hilbert space.
Let $\fff=\bigoplus_{n=0}^\infty \left(\otimes_{\rm{sym}}^n \LR\right)$ be the boson Fock space
over $\LR$, where the subscript means symmetrized tensor product. We denote the boson annihilation
and creation operators by $a(f)$ and $\add(f)$, $f,g\in\LR$, respectively, satisfying the canonical
commutation relations
\eq{ccr}
[a(f), \add(g)]=(\bar f, g), \quad [a(f),a(g)]=0=[\add(f),\add(g)].
\en
We use the informal expression $\ass(f)=\int \ass(k) f(k) dk$ for notational convenience. Consider
the Hilbert space
\eq{hilbertspace}
\hhh=\CC^2\otimes\fff.
\en
Denote by $d\Gamma(T)$ be the second quantization of a self-adjoint operator $T$ in $\LR$. The operator
on Fock space defined by
\eq{freehamiltonian}
\hf=d\Gamma(\omega)
\en
is the {free boson Hamiltonian} with dispersion relation
$
\omega(k)=|k|$.
The operator
\eq{freefield}
\field (\hat h)=\frac{1}{\sqrt 2}\int \left(\add(k)\hat h(-k)+a(k) \hat h(k)\right) dk,
\en
acting on Fock space is the {scalar field operator}, where $h\in\LR$ is a suitable form factor and
$\hat h$ is the Fourier transform of $h$. Denote by $\s_x,\s_y$ and $\s_z$ the $2\times 2$ Pauli
matrices given by
\eq{paulimatrices}
\s_x=\mmm 0 1 1 0,\ \ \ \s_y=\mmm 0 {-i} i 0 ,\ \ \ \s_z=\mmm 1 0 0 {-1}.
\en
With these components, the spin-boson Hamiltonian is defined by the linear operator
\eq{s4-1}
\SB=\eps \s_z\otimes\one +
\one\otimes \hf +\alpha \s_x\otimes \field (\hat h)
\en
on
$\hhh$, where $\alpha\in \RR$ is a  coupling constant and $\eps\geq0$ a parameter.

\subsection{A Feynman-Kac-type formula}
In this section we give a functional integral representation of $e^{-t\SB}$ by making use of a
Poisson point process and an infinite dimensional Ornstein-Uhlenbeck process. First we transform $\SB$
in a convenient form to study its spectrum in terms of path measures.

Recall that the rotation group in $\RR^3$ has an adjoint representation on $SU(2)$. Let $n\in \RR^3$
be a unit vector and $\theta\in [0,2\pi)$. Thus $e^{(i/2)\theta n\cdot \s}$ satisfies that
$e^{(i/2)\theta n\cdot \s} \s_\mu e^{-(i/2)\theta n\cdot \s}=(R\s)_\mu$, where $R$ denotes the $3\times 3$
matrix representing the rotation around $n$ with angle $\theta$, and $\s=(\s_x,\s_y,\s_z)$. In particular,
for $n=(0,1,0)$ and $\theta=\pi/2$, we have
\begin{align}
\label{angle1}
&e^{(i/2)\theta n\cdot \s} \s_x e^{-(i/2)\theta n\cdot \s}=\s_z,\\
&
\label{angle2}
e^{(i/2)\theta n\cdot \s} \s_z e^{-(i/2)\theta n\cdot \s}=-\s_x.
\end{align}
Let
\begin{align}
\label{unitary}
U=\exp\lk i\frac{\pi}{4}\s_y\rk\otimes \one = \frac{1}{\sqrt 2}\mmm 1 1 {-1} 1 \otimes \one
\end{align}
be a unitary operator on $\hhh$. By \kak{angle1} and \kak{angle2} $\SB$ transforms as
\eq{s4-2}
\PF=U \SB U^\ast=-\eps \s_x\otimes\one +\one\otimes \hf +\alpha \s_z\otimes \field (\hat h).
\en
Then $\PF$ is realized as
$$
\PF=\mmm {\hf+\alpha\field (\hat h)} {-\eps}{- \eps} {\hf-\alpha\field (\hat h)}.
$$
In particular, $\eps=0$ makes $\PF$ diagonal. If $\hat h/\sqrt\omega \in\LR$ and $h$ is real-valued, then
$\field (\hat h)$ is symmetric and infinitesimally small with respect to $\hf$, hence by the Kato-Rellich
theorem it follows that $\PF$ is a self-adjoint operator on $D(\hf)$ and bounded from below.

To construct the functional integral representation of the semigroup $e^{-t\PF}$, it is useful to introduce
a spin variable $\s\in \zz$, where $\zz=\{-1,+1\}$ is the additive group of order~2.
For $\Psi=\vvv{\Psi(+)\\ \Psi(-)}\in \hhh$, we have
\eq{s8}
\PF\Psi=\vvv{(\hf+\alpha\field (\hat h))\Psi(+)-\eps \Psi(-)\\
(\hf-\alpha\field  (\hat h))\Psi(-)-\eps \Psi(+)}.
\en
Thus we can transform $\PF$ on $\hhh$ to the operator $\tPF$ on
\eq{s9}
L^2(\zz;\fff)=\lkk f:\zz\rightarrow \fff\left | \|f\|_{L^2(\zz;\fff)}=
\sum_{\s\in\zz}\|f(\s)\|_\fff^2<\infty\right.\rkk
\en
by
\eq{s2}
(\tPF  \Psi)(\s) = \lk \hf +\alpha\s \field (\hat h)\rk  \Psi(\s)+\eps \Psi(-\s), \quad\s\in\zz.
\en
In what follows, we identify the Hilbert space $\hhh$ with $L^2(\zz;\fff)$ through
$$
\hhh \ni \vvv {{\Psi(+)}\\ \Psi({-})}\mapsto \Psi(\s)=
\lkk \begin{array}{ll} \Psi(+),&\s=+1,\\
\Psi(-),& \s=-1\end{array}\right.\in L^2(\zz;\fff),
$$
and instead of $\PF$ we consider $\tPF$, and use the notation $\PF$ for $\tPF$.

Let $(\Omega, \Sigma, P)$ be a probability space, and $\pro N$ be a two-sided Poisson process with
unit intensity on this space. We denote by $D=\{t\in\RR \,|\, N_{t+} \neq N_{t-}\}$ the set of jump
points, and define the integral with respect to this Poisson process by
$$
\int _{(s,t]} f(r, N_{r}) dN_r=
\sum_{\stackrel{r\in D}{r\in (s,t]}} f(r,N_r)
$$
for any predictable function $f$ (we refer to Appendix of  \cite{hl08} for details).
In particular,
we have for any continuous function $g$,
\eq{nt}
\int _{(s,t]} g(r, N_{r-}) dN_r=
\sum_{\stackrel{r\in D}{s< r\leq t}} g(r,N_{r-}).
\en
We  write
$\int _{s}^{t+}\cdots  dN_r$
for $\int _{(s,t]} \cdots dN_r$.
Note that $\int _{s}^{t+} g(r, N_{-r}) dN_r$ is right-continuous in $t$ and the integrand  $g(r, N_{-r})$
is left-continuous in $r$ and thus a predictable process. Define the random process
\eq{path}
\s_t=\s (-1)^{N_t},\quad\s\in\zz.
\en
This process describes the spin. Since our Poisson process is indexed by the real line, we summarize its
properties below.
\begin{proposition}\label{poisson}
The stochastic process $\pro N$ has the following properties:
\begin{enumerate}
\item
\emph{Independence:} The random variables $N_t$ and $N_s$ are independent for all $s\leq 0\leq t$, $s\neq t$.
\item
\emph{Markov property:} The stochastic processes $(N_t)_{t\geq 0}$ and $(N_{t})_{t\leq 0}$ are Markov processes
with respect to the natural filtrations $\nnn_t^+=\s(N_s, 0\leq s\leq t)$ and $\nnn_t^-=\s(N_s, t\leq s\leq 0)$,
respectively, i.e.,
$$\Ebb_P\lkkk N_{t+s}|\nnn_s^+\rkkk=\Ebb_P^{N_s}\lkkk N_t\rkkk,\qquad
\Ebb_P\lkkk N_{-t-s}|\nnn_{-s}^-\rkkk=\Ebb_P^{N_{-s}}\lkkk N_{-t}\rkkk.$$
\item
\emph{Reflection symmetry:} The random variables $N_t$ and $N_{-t}$ are identically distributed for all $t\in\RR$,
i.e., $\d
\Ebb_P[f(N_{-t})]=\Ebb_P[f(N_t)]=\sum_{n=0}^\infty f(n) \frac{|t|^n}{n!}e^{-|t|}$.
\item
\emph{Shift invariance:} The stochastic process $\s_t=\s (-1)^{N_t}$, $t\in\RR$, is shift invariant, i.e.,
$$\sum_{\s\in\zz}\Ebb_P\lkkk \prod_{j=0}^n f_j(\s_{t_j})\rkkk =\sum_{\s\in\zz}\Ebb_P\lkkk
\prod_{j=0}^n f_j(\s_{s+t_j})\rkkk,\quad s\in\RR.
$$
\end{enumerate}
\end{proposition}
\proof
The proof is a minor modification of \cite[Theorem 3.106]{lhb11} and it is omitted.

\qed
In the Schr\"odinger representation the boson Fock space $\fff$ can be realized as an $L^2$-space over
a probability space $(Q,\mu)$, and the field operator $\field (\hat f)$ with real-valued function $f\in\LR$
as a multiplication operator, which we will denote by $\phi(f)$. The identity function $\one$ on $Q$ corresponds
to the Fock vacuum $\ob$ in $\fff$.

Next we introduce the random process describing the free boson field $\hf$. Let $(Q_{\rm E},\mu_{\rm E})$ be
a probability space associated with the Euclidean quantum field (for details see \cite[p.268-p.276]{lhb11}).
The Hilbert spaces $L^2(Q_{\rm E})$ and $L^2(Q)$ are related through the family of isometries $\{j_s\}_{s\in\RR}$
from $\LR$ to $\LRT$ defined by
$$
\d  \widehat {j_s f}(k,k_0)= \frac{e^{-itk_0}}{\sqrt{\pi}} \sqrt{\frac{\omega(k)}{|k_0|^2+\omega(k)^2}}\hat f(k),
$$
where $\hat f$ denotes the Fourier transform of $f$. Let $\Phi_{\rm E}(j_sf)$ be a Gaussian random variable on
$(Q_{\rm E},\mu_{\rm E})$ indexed by $j_sf\in\LRT$ with mean zero and covariance
$$
\Ebb_{\mu_{\rm E}}[\Phi_{\rm E}(j_sf)\Phi_{\rm E}(j_t g)]=
\half\int_\BR e^{-|s-t|\omega(k)} \ov{\hat f(k)}\hat g(k) dk.
$$
Also, let $\{J_s\}_{s\in\RR}$ be the family of isometries from $L^2(Q)$ to $L^2(Q_{\rm E})$ defined by
$$
J_s\wick{\phi(f_1)\cdots \phi(f_n)} = \wick{\Phi_{\rm E}(j_s f_1)\cdots \Phi_{\rm E}(j_s f_n)},
$$
where $\wick{X}$ denotes Wick product of $X$.
Then we derive that
\eq{jj}
(J_s\Phi, J_t\Psi)_{L^2(Q_{\rm E})}=
(\Phi, e^{-|t-s|\hf}\Psi)_{L^2(Q)}.
\en

In \cite{hl08} by making use of the process $\pro \s$ a functional integral representation of the Pauli-Fierz
model with spin $\han$ in non-relativistic quantum electrodynamics was obtained. By a suitable modification we
can also construct the functional integral representation of $e^{-t\PF}$. In fact, the construction in the
spin-boson case becomes simpler than in the case of the Pauli-Fierz model, see also \cite[Remark 6.3]{hl08}
and \cite[p.424-p.454]{lhb11}. We have the following Feynman-Kac-type formula for the spin-boson Hamiltonian.

We identify $\hhh$ as
\eq{id}
\hhh\cong L^2(\zz; L^2(Q))\cong L^2(\zz\times Q).
\en
\bp{main}
Let $\Phi,\Psi\in\hhh$ and $h\in\LR$ be real-valued.
Then
\begin{align}
\label{hl}
&
(\eps\not=0)\quad
(\Phi, e^{-t\PF}\Psi)_\hhh = e^t \sum_{\s\in\zz} \EE  \EEE \left[\ov{J_0\Phi(\s_0)}
e^{-\alpha \Phi_{\rm E}\lk \int_0^t \s_s j_s h ds\rk } \eps^{N_t}J_t\Psi(\s_t) \right]\\
&\label{hl1}
(\eps=0) \quad (\Phi, e^{-t\PF}\Psi)_\hhh = e^t \sum_{\s\in\zz} \EEE \lkkk
\ov{J_0\Phi(\s)} e^{-\alpha \Phi_{\rm E}\lk \s \int_0^t  j_s h ds\rk } J_t\Psi(\s)\rkkk.
\end{align}
\ep
\proof
Let $\eps\not=0$. We see by \kak{s2} that
$$
\PF \Psi(\s)=(\hf+\alpha\phi(h))\Psi(\s)- e^{\log\eps}\Psi(-\s),\quad \s\in\zz.
$$
Then by \cite[Theorem 4.11]{hl08} we have
\eqn
(\Phi, e^{-t\PF}\Psi)_\hhh = e^t\sum_{\s\in\zz} \EE \EEE \left[ \ov{J_0\Phi(\s_0)}
e^{-\alpha\int_0^t\s_s\Phi_{\rm E}(j_sf) ds+\int_0^{t+} \log \eps dN_s} J_t\Psi(\s_t)\right].
\enn
Since $\int_0^{t+} \log\eps  dN_s= N_t \log \eps$, \kak{hl} follows.
Next consider the case $\eps=0$. As $\eps\rightarrow 0$ on both sides of \kak{hl}
the integrands over $\{N_t\geq1\}$ vanish while those on $\{N_t=0\}$ are non-vanishing. Moreover,
note that $N_s=0$, $s\leq t$, on  $\{N_t=0\}$. Hence \kak{hl1} is obtained by taking the limit
\eqn
\lim_{\eps\rightarrow 0}(\Phi, e^{-t\PF}\Psi)_\hhh
&=&
\lim_{\eps\rightarrow 0} e^t \sum_{\s\in\zz} \EE  \EEE \left[ \ov{J_0\Phi(\s_0)}
e^{-\alpha \Phi_{\rm E}\lk \int_0^t \s_s j_s h ds\rk} \eps^{N_t}J_t\Psi(\s_t) \right]\\
&=&
e^t \sum_{\s\in\zz} \EEE\lkkk \ov{J_0\Phi(\s)}e^{-\alpha \Phi_{\rm E}\lk  \s \int_0^t  j_s h ds\rk}
J_t\Psi(\s)\rkkk.
\enn
\qed
\br{rep1}
{\rm
By the Feynman-Kac formula (Proposition \ref{main}) we see that
\eq{rep}
e^{-t\PF}\Phi(\s)=e^t \Ebb_P\lkkk
J_0^\ast e^{\Phi_{\rm E}(-\alpha\int_0^t \s_rj_r h dr)}J_t \Phi(\s_t) \rkkk
\en
for every $\s\in\zz$.
}
\er

Denote $\onee=\one_{\LZ}\otimes\one_{L^2(Q)}$. Using the above proposition we can compute the vacuum
expectation of the semigroup $e^{-t\PF}$, which plays an important role in this paper.
\bc{vacuumexpectation}
Let $h\in\LR$ be a real-valued function. Then for every $t>0$ it follows that
\eq{ve}
(\onee, e^{-t\PF}\onee)=
e^t\sum_{\s\in \zz} \Ebb_P\lkkk \eps^{N_t}e^{\frac{\alpha^2}{2}\int_0^t dr \int_0^t W(N_r-N_s,r-s)ds}\rkkk,
\en
where the pair interaction potential $W$ is given by
\eq{W}
W(x,s)=\frac{(-1)^x}{2}  \int_\BR e^{-|s|\omega(k)} {|\hat h(k)|^2}dk.
\en
\ec
\proof
By Proposition \ref{main} we have
$$
(\onee, e^{-t\PF}\onee)=
e^t\sum_{\s\in \zz} \Ebb_P\lkkk \eps^{N_t}e^{\frac{\alpha^2}{2}\int_0^t dr \int_0^t W(N_r+N_s,r-s)ds} \rkkk.
$$
Since $W(N_r+N_s,r-s)=W(N_r-N_s,r-s)$, the corollary follows.
\qed
\noindent
Note that equality \kak{W} gives the interaction potential
\eq{hosono}
W(N_r-N_s,r-s)=\half \s_r \s_s  \int_\BR e^{-|r-s|\omega(k)} {|\hat h(k)|^2} dk
\en
of an infinite range Ising-model on the real line instead of a lattice.

\subsection{Parity symmetry}
It is a known fact that $\SB$ has a parity symmetry. Let
\eq{parity1}
P=\s_z\otimes (-1)^N,
\en
where $N=d\Gamma(\one)$ denotes the number operator in $\fff$. From $\spec(\s_z)=\{-1,1\}$ and
$\spec(N)=\{0,1,2,...\}$ it follows that $\spec(P)=\{-1,1\}$. We identify $\hhh$ with $\fff_\uparrow
\oplus \fff_\downarrow $, where $\fff_\uparrow $ and $\fff_\downarrow $ are identical copies of $\fff$.
Then each Pauli matrix $\s_X=\mmm a b c d $ acts  as
$$\s_X\vvv{\Psi(+) \\
\Psi(-)}=\vvv  {a\Psi(+)+b\Psi(-)\\
c\Psi(+)+d\Psi(-)}$$
for
$\vvv{\Psi(+) \\ \Psi(-) }\in \fff_\uparrow  \oplus \fff_\downarrow  $.
Furthermore, $\fff=\oplus_{n}^\infty\fff_n$ can be decomposed  as $\fff_{\rm e}\oplus \fff_{\rm o}$,
where  $\fff_{\rm e}$ and $\fff_{\rm o}$ denote respectively the subspaces of $\fff$ consisting of
even and odd numbers of bosons, i.e., $\fff_{\rm e}=\oplus_{m=0}^\infty\fff_{2m}$ and $\fff_{\rm o}=
\oplus_{m=0}^\infty\fff_{2m+1}$. The projections from $\fff$ to $\fff_{\rm e}$ and $\fff_{\rm o}$ are
denoted by $P_{\rm e}$ and $P_{\rm o}$, respectively. Let $\hhh_+=P_{\rm e}\fff_\uparrow \oplus
P_{\rm o} \fff_\downarrow $ and $\hhh_-= P_{\rm o}\fff_\uparrow \oplus P_{\rm e} \fff_\downarrow $
be subspaces of $\fff_\uparrow \oplus\fff_\downarrow $.
\bl
{parity}
The following properties hold:
\begin{itemize}
\item[(1)]
The Hilbert space  $\hhh$ can be identified with $\hhh_+\oplus\hhh_-$
by the correspondence
$$
\fff_\uparrow  \oplus\fff_\downarrow  \ni \vvv{\Psi(+)\\ \Psi(-)}\mapsto
\vvv{\Psi_{\rm e}(+)\\ \Psi_{\rm o}(-)}\oplus \vvv{\Psi_{\rm o}(+)\\ \Psi_{\rm e}(-)}\in \hhh_+\oplus\hhh_-,
$$
where $\Psi_{\rm e}(\pm)=P_{\rm e}\Psi(\pm)$ and $\Psi_{\rm o}(\pm)=P_{\rm o}\Psi(\pm)$.
\item[(2)]
It follows that $[\SB,P]=0$.
\item[(3)]
$\hhh_\pm$ is the eigenspace associated with eigenvalue $\pm 1$  of $P$.
\item[(4)]
$\SB$ can be decomposed as $\SB=\SB\lceil_{\hhh_+}\oplus \SB\lceil_{\hhh_-}$.
\end{itemize}
\el
\proof
(1) and (2) are  straightforward. Let $\Psi=\vvv{\Psi_{\rm e}(+)\\
\Psi_{\rm o}(-)}\in \hhh_+$. Then
$P\Psi=\Psi$ follows by a direct calculation. Thus $\Psi$ is eigenvector of $P$ with eigenvalue
$+1$. Similarly it follows that $\hhh_-$ is the eigenspace associated with eigenvalue $-1$ of $P$,
and (3) also follows. (4) is obtained by a combination of (1), (2)  and (3).
\qed

Finally, for later use we show some related spin-flip properties.
\bl{flip}
We have the following properties:
\begin{itemize}
\item[(1)]
$(\Psi, \s_x\Phi)=0$ and $(U^\ast \Psi, \s_zU^\ast  \Phi)=0$ for any $\Psi,\Phi\in\hhh_\pm$.
\item[(2)]
$(\Psi, \phi(f)\Phi)=0$ and $(U^\ast \Psi, \phi(f)U^\ast  \Phi)=0$ for any $\Psi,\Phi\in\hhh_\pm$.
\end{itemize}
\el
\proof
(1) It is straightforward to show that $\s_x$ is a spin-flip transform, i.e.,
$$
\s_x\hhh_\pm \subset \hhh_\mp,
$$
which gives the first statement. The second statement follows by observing that $U \s_z U^\ast =\s_x$.
To obtain (2) it is again straightforward to show  that
$$
\phi(f)\lk \hhh_\pm \cap D(\phi(f))\rk \subset \hhh_\mp.
$$
The second part follows by $U \phi(f)  U^\ast =\phi(f) $.
\qed

\section{Ground state of the spin-boson}
\subsection{Positivity improving semigroup}
In the remainder of this paper we assume that $h\in\LR$ is  real-valued.

Let $E=\inf\spec(\PF)$. We estimate the dimension of ${\rm Ker} (\PF-E)$ for $\eps\not=0$.
\bc{positivityimproving}
Assume that $\eps\not=0$. Then $e^{-t\PF }$, $t>0$, is a positivity improving semigroup on
$L^2(\zz\times Q)$, i.e., $(\Psi, e^{-t\PF }\Phi)>0$ for $\Psi, \Phi \geq 0$ such that $\Psi\not\equiv0
\not\equiv \Phi$.
\ec
\proof
The proof simplifies an argument in \cite{h9}.
It is trivial that $(\Psi, e^{-t\PF }\Phi)\geq0$,
thus it suffices showing that $(\Psi, e^{-t\PF }\Phi)\ne 0$. Suppose the contrary. Then
we have
$$
\sum_{\s\in\zz} \Ebb_P   \left[\lk J_0\Psi(\s_0), e^{\Phi_{\rm E}\lk-\alpha\int_0^t \s_s j_sh ds\rk}
{\eps}^{N_t}J_t\Phi(\s_t)\rk_{L^2(Q_{\rm E})} \right]=0.
$$
Since $J_t$ is positivity preserving and $e^{\Phi_{\rm E}\lk-\alpha\int_0^t \s_s
j_s h ds\rk}$ is positive,
for $\s\in\zz$ we have
$$
\Ebb_P  \left[\lk J_0\Psi(\s_0), e^{\Phi_{\rm E} \lk -\alpha\int_0^t \s_s j_s h  ds \rk}
{\eps}^{N_t}J_t\Phi(\s_t)\rk_{L^2(Q_{\rm E})}\right]=0,
$$
which implies that ${\rm supp} J_0\Psi(\s_0)\cap {\rm supp} J_t\Phi(\s_t)=\emptyset$ a.s. Hence
$0=(J_0\Psi(\s_0), J_t\Phi(\s_t))=(\Psi(\s_0), e^{-t\hf} \Phi(\s_t))$. Since $e^{-t\hf}$ is positivity
improving, $\Psi(\s_0)\equiv 0$ or $\Phi(\s_t)\equiv 0$. This contradicts that $\Psi\not\equiv 0$ and
$\Phi\not\equiv 0$, and the claim follows.
\qed

\subsection{Existence and uniqueness of ground state}
\subsubsection{The case of $\eps=0$}
Whenever $\eps=0$ the Hamiltonian $\PF$ is diagonal, i.e., we have
$$
\PF=\mmm{\hf+\alpha\phi(h)} 0 0 {\hf-\alpha\phi(h)}.
$$
It is known that $\hf+\alpha\phi(h)$, the Hamiltonian of the van Hove model, has a unique ground state
if and only if $\hat h/\omega\in\LR$, see e.g. \cite{hir06}, which implies that $\PF$ has a two-fold
degenerate ground state if and only if  $\hat h/\omega\in \LR$.

\subsubsection{The case of $\eps\not=0$}
Next we consider the case of $\eps\ne 0$.
 Write
\eq{groundstateapp}
\Phi_T=e^{-T(\PF-E)}\onee,\quad T\geq0,
\en
and
\eq{gamma}
\gamma(T)=\frac{(\onee, \Phi_T)^2}{\|\Phi_T\|^2}=\frac{(\onee, e^{T\PF}\onee)^2}{(\onee, e^{-2T\PF}\onee)}.
\en
A known criterion  of existence of a ground state is \cite[Proposition 6.8]{lhb11}.
\begin{proposition}
\label{criter}
A ground state of $\PF$ exists if and only if $\d \lim_{T\to\infty}\gamma(T)>0$.
\end{proposition}
By Corollary \ref{vacuumexpectation}
 we have
\begin{align}
\|\Phi_T \|^2&=e^{2TE}\sum_{\s\in\zz} \Ebb_P
\lkkk \eps^{N_T}e^{\frac{\alpha^2}{2}\int_{-T}^T dt\int_{-T}^T W(N_t-N_s, t-s) ds}\rkkk,\\
\label{reflection}
(\onee, \Phi_T)
&=
e^{TE}\sum_{\s\in\zz} \Ebb_P \lkkk
\eps^{N_T}
e^{\frac{\alpha^2}{2}\int_0^T dt\int_0^T W(N_t-N_s, t-s) ds} \rkkk\non\\
&=
e^{TE}\sum_{\s\in\zz} \Ebb_P
\lkkk \eps^{N_T}
e^{\frac{\alpha^2}{2}\int_{-T}^0 dt\int_{-T}^0 W(N_t-N_s, t-s) ds } \rkkk.
\end{align}
The second identity on \kak{reflection} is derived from the reflection symmetry in Proposition~\ref{poisson}.
Note that
\eq{bound}
\left| \int_{-T}^ 0 dt \int_0^T W(N_t-N_s, t-s) ds  \right|\leq \frac{1}{2}\left\|\frac{\hat h}{\omega}\right\|^2
\en
uniformly in $T$ and in the paths.
\bt{existence}
If $\hat h/\omega\in\LR$, then $\PF$ has a ground state and it is unique.
\et
\proof
We write $\d \int_{-T}^T\!\!\! dt\int_{-T}^T W ds =
\int_{-T}^0 \!\!\! dt\int_{-T}^0 W ds + \int_0^T \!\!\!dt\int_0^T W ds +
2\int_{-T}^0 \!\!\!dt\int_0^T W ds$, and by \kak{bound} obtain
\begin{align}
\|\Phi_T\|^2 \leq
e^{2TE} \sum_{\s\in\zz}\Ebb_P\lkkk \eps^{N_T}e^{\frac{\alpha^2}{2}(\int_{-T}^0 dt\int_{-T}^0 ds W(N_t-N_s, t-s)  +
\int_0^T dt\int_0^T ds W(N_t-N_s, t-s)  +\|\hat h/\omega\|^2)} \rkkk.
\end{align}
By the independence of $N_t$ and $N_{-s}$, and reflection symmetry of the paths we furthermore obtain that
\begin{align*}
\|\Phi_T\|^2 &\leq e^{2TE}
\sum_{\s\in\zz} \lk \Ebb_P\lkkk \eps^{N_T}e^{\frac{\alpha^2}{2} \int_0^T dt\int_0^T ds W(N_t-N_s, t-s)  }
\rkkk\rk^2 e^{{\frac{\alpha^2}{2}}\|\hat h/\omega\|^2}\\
&\leq
\lk e^{TE} \sum_{\s\in\zz} \Ebb_P\lkkk \eps^{N_T}e^{\frac{\alpha^2}{2} \int_0^T dt\int_0^T ds W(N_t-N_s, t-s) }
\rkkk\rk^2 e^{{\frac{\alpha^2}{2}}\|\hat h/\omega\|^2}\\
&=(\onee, \Phi_T)^2
e^{{\frac{\alpha^2}{2}}\|\hat h/\omega\|^2}.
\end{align*}
Hence $\gamma(T)\geq e^{-{\frac{\alpha^2}{2}}\|\hat h/\omega\|^2}$ and a ground state $\gr$ of $\PF$ exists.
By Corollary \ref{positivityimproving} $\gr$ is strictly positive as a vector in $L^2(\zz\times Q)$,
in particular, it is unique.
\qed

We note that the condition $\hat h/\omega\in\LR$ in Theorem \ref{existence} is not a mere technicality.
This condition will play an essential role throughout below in the definition of a Gibbs measure and the
analysis of the ground state properties, see also Section \ref{L2} below.

By Theorem \ref{existence} it follows that $\SB$ also has a unique ground state.  As seen above, $\hhh$ can be
decomposed as $\hhh=\hhh_+\oplus\hhh_-$ and $\PF$ can be reduced by $\hhh_\pm$.
\bc{flip2}
Let $\grsb $ be the ground state of $\SB$. Then  $\grsb  \in\hhh_{-}$.
\ec
\proof
Let $U$ be as in \kak{unitary}. Notice that $\grsb =U^\ast \gr$, and thus
$$
\d \grsb =s-\lim_{T\to\infty} \frac{U^\ast e^{-T\PF}\onee}{\|U^\ast e^{-T\PF}\onee\|} =
s-\lim_{T\to\infty}  \frac{e^{-T\SB} U^\ast \onee}{\|e^{-T\SB} U^\ast \onee\|}.
$$
The function $\onee\in \LZ\otimes L^2(Q)$ corresponds to $\vvv {\Omega \\\Omega }\in \fff_\uparrow
\oplus\fff_\downarrow$ and
$$
U^\ast \onee =\half \mmm 1 {-1} 1 1 \vvv {\Omega\\ \Omega}= \vvv{0\\\Omega}\in\hhh_-.
$$
Hence by the parity symmetry of $\SB$ we have
$$
Pe^{-T\SB} U^\ast \onee =e^{-T\SB} P U^\ast \onee = -e^{-T\SB} U^\ast \onee
$$
and thus $e^{-T\SB} U^\ast \onee \in \hhh_{-}$. This implies that $\grsb \in\hhh_-$.
\qed

\begin{remark}
{\rm
By Corollary \ref{positivityimproving} the ground state $\gr$ of $\PF$ overlaps with  the non-negative vector
$\d \rho(\s,\phi)=\lkk\begin{array}{ll}
1,& \s=+1\\
0,&\s=-1
\end{array}
\right.
$
in  $L^2(\zz\times Q)$. Hence $(\gr, \rho)_{L^2(\zz\times Q)}\not=0$
and
\begin{align}
\is(\PF) = -\lim_{\beta\rightarrow \infty}\frac{1}{\beta}\log (\rho, e^{-\beta\PF} \rho)
= -\lim_{\beta\rightarrow \infty}\frac{1}{\beta}\log e^\beta \Ebb_P
\left[\eps^{N_t}e^{\frac{\alpha^2}{2} \int_0^\beta dt \int_0^\beta ds W}\right].
\end{align}
The expression at the right hand side above was also obtained in \cite{hir99,abd12}.
}
\end{remark}

\section{Path measure associated with the ground state}
\subsection{$\zz$-valued paths}
\label{s31}
In Sections \ref{s31}-\ref{s32} we set $\eps=1$ for simplicity.

Let $\ms X=D(\RR;\zz)$ be the space of c\`adl\`ag paths with values in $\zz$, and $\ms G$ the $\s$-field
generated by cylinder sets. Thus $\s_ \cdot :(\Omega, \Sigma, P)\to (\ms X, \ms G)$ is an $\ms X$-valued
random variable. We denote its image measure by $\P^\s$, i.e.,  $\P^\s(A)=\s_ \cdot^{-1}(A)$ for $A\in
\ms G$, and the coordinate process by $\pro X$, i.e., $X_t(\omega)=\omega(t)$ for $\omega \in \ms X$.
Hence Proposition~\ref{main} can be reformulated in terms of $\pro X$ as
\begin{eqnarray}
(\Phi, e^{-t\PF}\Psi)_\hhh = e^t \sum_{\s\in\zz} \Ebb_\P^\s   \EEE \left[\ov{J_0\Phi(X_ 0)}
e^{-\alpha \Phi_{\rm E}\lk \int_0^t X_ s j_s h ds\rk }  J_t\Psi(X_ t) \right].
\end{eqnarray}
Here $\Ebb_{\P^\s}=\Ebb_\P^\s$ so that $\Ebb_\P^\s[X_ 0=\s]=1$.
Then \kak{rep} can be converted to the form
\eq{rep2}
e^{-t\PF}\Phi(\s)=e^t  \Ebb_\P^\s \lkkk
J_0^\ast e^{\Phi_{\rm E}(-\alpha\int_0^t X_rj_r h dr)}J_t \Phi(\s_t) \rkkk
\en
for every $\s\in\zz$.
\bl{shift}
For every  $s,t\in\RR$ it follows that
\begin{eqnarray}
(\Phi, e^{-t\PF}\Psi)_\hhh = e^t \sum_{\s\in\zz} \Ebb_\P^\s   \EEE \left[\ov{J_s\Phi(X_ s)}
e^{-\alpha \Phi_{\rm E}\lk \int_s^{s+t} X_ r j_r h dr\rk }  J_{s+t}\Psi(X_ {s+t}) \right].
\end{eqnarray}
\el
\proof
By the Trotter product formula $\d (\Phi, e^{-t\PF}\Psi)_\hhh =
\lim_{n\to\infty}(\Phi, (e^{-\frac{t}{n}H_0}e^{-\frac{t}{n}\hf})^n \Psi)$,
and using the fact that $e^{-|t-s|\hf}=J_t^\ast J_s$ we have
$$
(\Phi, e^{-t\PF}\Psi)_\hhh = e^t \sum_{\s\in\zz} \Ebb_P \EEE \left[\ov{J_s\Phi(\s_0)}
e^{-\alpha \Phi_{\rm E}\lk \int_0^ t  \s_ r j_{s+r} h dr\rk }  J_{s+t}\Psi(\s_ t) \right].
$$
By the shift invariance of $\pro N$ stated in Proposition \ref{poisson} we have
$$
(\Phi, e^{-t\PF}\Psi)_\hhh = e^t \sum_{\s\in\zz} \Ebb_P \EEE \left[\ov{J_s\Phi(\s_s)}
e^{-\alpha \Phi_{\rm E}\lk \int_0^ t  \s_{s+ r} j_{s+r} h dr\rk }  J_{s+t}\Psi(\s_ {s+t}) \right].
$$
Hence the lemma follows.
\qed

Let $(\zz, \ms B)$ be a measurable space with $\s$-field  $\ms B=\{\emptyset, \{-1\},\{+1\},\zz\}$.
For later use we show a functional integral representation of Euclidean Green functions of the type
$(\Phi, \one_{A_0}e^{-(t_1-t_0)\PF} \one_{A_1} e^{-(t_2-t_1)\PF}\cdots e^{-(t_n-t_{n-1})\PF}\one_{A_n}\Psi)$,
where $-\infty<t_0\leq\ldots\leq t_n<\infty$ and $A_0,...,A_n\in \ms B$. We see that the operator
\eq{qboundedoperator}
Q_{[S,T]}={J_S^\ast e^{\Phi_{\rm E}
(-\alpha\int_S^T X_ s j_s h ds)} J_T}:L^2(Q)\to L^2(Q)
\en
is bounded. In fact, we have that
\eq{44}
\|Q_{[S,T]}\|_{L^2(Q)\to L^2(Q)} \leq \|Q_{[S,T]}\|_{L^1(Q)} \leq e^{\frac{\alpha^2}{4}\|\int_S^T X_ s j_s h ds\|^2},
\en
which was shown in e.g. \cite[Corollary 4.4]{hl08}.
\bc{greenfunction}
Let $-\infty<t_0\leq\ldots\leq t_n<\infty$ and $A_0,...,A_n\in \ms B$. Then
\begin{eqnarray}
\lefteqn{(\Phi, \one_{A_0}e^{-(t_1-t_0)\PF} \one_{A_1} e^{-(t_2-t_1)\PF}\cdots
e^{-(t_n-t_{n-1})\PF}\one_{A_n}\Psi)}\non \\
&&
\label{greenfunctions}
= e^{t_n-t_0} \sum_{\s\in\zz}\Ebb_{\P}^\s\Ebb_{\mu_{\rm E}} \lkkk \lk  \prod_{j=0}^n \one_{A_j}(X_ {t_j})\rk
\ov{\Phi(X_{t_0})} Q_{[t_0,t_n]}\Psi(X_ {t_n})\rkkk.
\end{eqnarray}
\ec
\proof
This is proven by using the Markov properties of both the Poisson process and the Euclidean field, and
\kak{rep}.
Denote by ${\ms N}_s=\s(N_r,0\leq r\leq s)$ the natural filtration of the Poisson process $\pro N$.
The Markov property of $\pro N $ and \kak{rep} yield that
\begin{align*}
&
\hspace{-1cm}
\lk e^{-s\PF}\one_A e^{-t\PF} \Phi\rk (\s)\\
&=
e^{s+t} \Ebb_P\lkkk J_0^\ast e^{\Phi_{\rm E}(-\alpha\int_0^s \s_rj_r h dr)}J_s \one_A(\s_s)
\Ebb_P \lkkk \left.
J_0^\ast e^{-\alpha\Phi_{\rm E}(\int_0^t\s_{r+s}j_rh dr)} J_t
\Phi(\s_{t+s})\right|{\ms N}_s \rkkk \rkkk\\
&=
e^{s+t} \Ebb_P \lkkk J_0^\ast e^{\Phi_{\rm E}(-\alpha\int_0^s \s_rj_r h dr)}J_s \one_A(\s_s) J_0^\ast
e^{-\alpha\Phi_{\rm E}(\int_0^t\s_{r+s}j_rh dr)}J_t \Phi(\s_{t+s}) \rkkk\\
&=
e^{s+t} \Ebb_P\lkkk J_0^\ast e^{-\alpha\Phi_{\rm E}(\int_0^s \s_r j_r  h dr)} J_s \one_A(\s_s)
J_0^\ast e^{-\alpha\Phi_{\rm E}(\int_0^t\s_{r+s}j_rh dr)}J_t \Phi(\s_{t+s}) \rkkk.
\end{align*}
Since $J_0=U_{-s} J_s$, where $U_s:L^2(Q_{\rm E})\to L^2(Q_{\rm E})$  is the shift operator defined
by $U_s\Phi_{\rm E}(j_{t_1}f_1)\cdots \Phi_{\rm E}(j_{t_n}f_n)=\Phi_{\rm E}(j_{t_1+s}f_1)\cdots
\Phi_{\rm E}(j_{t_n+s}f_n)$, we obtain
\begin{align}
&
\hspace{-0.5cm} \lk e^{-s\PF}\one_A e^{-t\PF} \Phi\rk (\s)\non \\
&=
e^{s+t} \Ebb_P \lkkk J_0^\ast e^{-\alpha\Phi_{\rm E}(\int_0^s j_r \s_r h dr)} J_sJ_s^\ast U_s
\one_A(\s_s) e^{-\alpha\Phi_{\rm E}(\int_0^t\s_{r+s}j_rh dr)}J_t \Phi(\s_{t+s}) \rkkk\non \\
&\label{markov} =
e^{s+t} \Ebb_P \lkkk J_0^\ast e^{-\alpha\Phi_{\rm E}(\int_0^s j_r \s_r h dr)} J_sJ_s^\ast
\one_A(\s_s)
e^{-\alpha\Phi_{\rm E}(\int_0^t \s_{r+s} j_{r+s}h dr)}J_{t+s} \Phi(\s_{t+s}) \rkkk.
\end{align}
Furthermore, by the Markov property of the Euclidean field we can remove
the projection $J_s J_s^\ast$ in \kak{markov}
and obtain
\begin{align*}
\lk e^{-s\PF}\one_A e^{-t\PF} \Phi\rk (\s) =
e^{s+t} \Ebb_P \lkkk J_0^\ast e^{-\alpha\Phi_{\rm E}(\int_0^{s+t} \s_r j_r  h dr)}\one_A(\s_s)
J_{t+s}\Phi(\s_{t+s}) \rkkk.
\end{align*}
Hence in terms of $\pro X$ we have
\begin{align*}
&
(\Phi, \one_{A_0}e^{-s\PF}\one_{A_1}e^{-t\PF}\one_{A_2}\Psi)\\
&=
e^{s+t}\sum_{\s\in\zz} \Ebb_\P^\s
\lkkk\! \one_{A_0}(X_0)\one_{A_1}(X_s)\one_{A_2}(X_{s+t})
\lk J_0 {\Phi(X_0)}, e^{-\alpha\Phi_{\rm E}(\int_0^{s+t}X_r j_r  h dr)}J_{t+s}
\Psi(X_{t+s})\rk\! \rkkk.
\end{align*}
Repeating this procedure, we have
\begin{eqnarray}
\lefteqn{(\Phi, \one_{A_0}e^{-(t_1-t_0)\PF} \one_{A_1} e^{-(t_2-t_1)\PF}\cdots
e^{-(t_n-t_{n-1})\PF}\one_{A_n}\Psi)}\non \\
&&
\label{greenfunctions11}
= e^{t_n-t_0} \sum_{\s\in\zz}\Ebb_{\P}^\s\Ebb_{\mu_{\rm E}} \lkkk \lk  \prod_{j=0}^n \one_{A_j}(X_ {t_j-t_0})\rk
\ov{\Phi(X_0)} Q_{[0,t_n-t_0]}\Psi(X_ {t_n-t_0})\rkkk.
\end{eqnarray}
By the shift invariance of $\s_t$  (Proposition \ref{poisson}) we complete the proof.
\qed

\bc{greenfunction2}
Let $-\infty<t_0\leq t_1\leq\ldots\leq t_n<\infty$ and $A_0,...,A_n\in \ms B$. Then
\begin{eqnarray}
\hspace{-1cm} \lefteqn{
(\one_{A_0},e^{-(t_1-t_0)\PF} \one_{A_1} e^{-(t_2-t_1)\PF}\cdots
e^{-(t_n-t_{n-1})\PF}\one_{A_n})}\non \\
&&
\label{greenvacuum}
=
e^{t_n-t_0}\sum_{\s\in\zz}\Ebb_{\P}^\s \lkkk e^{\frac{\alpha^2}{2}\int_{t_0}^{t_n} dt
\int_{t_0}^{t_n} ds W(X_s,X_t, t-s)}\prod_{j=0}^n \one_{A_j}(X_ {t_j})\rkkk,
\end{eqnarray}
where
$\d W(x,y,t)=\frac{xy}{2}\int_\BR e^{-|t|\omega(k)}{\hat h(k)^2} dk$.
\ec
\proof
By Corollary \ref{greenfunction}  we have
$$
\mbox{LHS} \, \kak{greenvacuum}
= e^{t_n-t_0} \sum_{\s\in\zz}\Ebb_{\P}^\s \lkkk\Ebb_{\mu_{\rm E}}\lkkk Q_{[t_0,t_n]}\rkkk
\prod_{j=0}^n \one_{A_j}(X_ {t_j})\rkkk.
$$
Hence the corollary follows.
\qed

\subsection{Local weak convergence}
\label{s32}
In this section we make the assumption that $\hat h/\omega\in\LR$, so that there is a unique ground
state $\gr \in \hhh$. Let ${\ms G}_{[-T,T]}=\sigma(X_ t, t\in [-T,T])$ be the family of sub-$\s$-fields
of $\ms G$
and
$${\cal G}=\bigcup_{T\geq0}{\ms G}_{[-T,T]}.$$
Let $\ov{\cal G}=\s(\cal G)$.
Define the probability measure $\mu_T$ on $(\ms X, \ov{\cal G})$ by
\eq{measure}
\mu_T(A)=\frac{e^{2T}}{Z_T} \sum_{\s\in\zz} \Ebb_\P^\s\lkkk \one_A
e^{\frac{\alpha^2}{2}\int_{-T}^T dt\int_{-T}^T ds W(X_t,X_s, t-s) }\rkkk, \quad A\in \ov{\cal G},
\en
where $Z_T$ is the normalizing constant such that $\mu_T (\ms X)=1$.  This probability measure is a Gibbs measure for the pair interaction potential $W$, indexed by the bounded intervals
$[-T,T]$ (see the next section for further details). In this section we show convergence of $\mu_T$ to a probability measure $\mu_\infty$ in a specific sense when $T\to\infty$.
\begin{definition}
\rm{
Let $\mu_\infty$ be a probability measure on $(\ms X,\ov{\cal G})$, and $\seq T \subset \RR$ be any unbounded increasing sequence of positive numbers. The sequence of probability measures $(\mu_{T_n})_{n\in \mathbb{N}}$ is said to converge to the probability measure $\mu_\infty$ in \emph{local weak topology} whenever
$\limn |\mu_{T_n}(A) - \mu_\infty(A)|=0$ for all $A\in \ms G_{[-t, t]}$ and $t\geq 0$.
}
\end{definition}
By the above definition it is seen that whenever $\mu_T\to\mu_\infty$ in local weak sense, we have that
\eq{lwc}
\lim_{T\to\infty} \Ebb_{\mu_T}[f]=\Ebb_{\mu_\infty}[f]
\en
 for any bounded $\ms G_{[-t,t]}$-measurable function $f$.

Next we define the finite dimensional distributions indexed by $\La=\{t_0,\ldots,t_n\}\subset [-T,T]$ with $t_0\leq\ldots\leq t_n$.
Let
\eq{finite}
\mu_T^\La (A_0\times\cdots\times A_n)= \frac{e^{2T}}{Z_T} \sum_{\s\in\zz} \Ebb_\P^\s\lkkk \lk
\prod_{j=0}^n \one _{A_{j}}(X_ {t_j}) \rk
e^{\frac{\alpha^2}{2}\int_{-T}^T dt\int_{-T}^T ds W(X_t,X_s, t-s) }\rkkk
\en
be a probability measure  on $(\zz^\La, \ms B^\La)$, where $\zz^\Lambda=\times_{j=1}^n \zz^{t_j} $
and $\ms B^\Lambda=\times_{j=1}^n \ms B^{t_j}$ for $\Lambda=\{t_1,...,t_n\}$, and $\zz^{t_j}$ and
$\ms B^{t_j}$ are copies of $\zz$ and $\ms B$, respectively.
Clearly, ${\cal G}$ is a finitely additive family of sets.
Define an additive set function on $(\ms X, {\cal G})$
by
\eq{setfunction}
\mu(A)=e^{2Et}e^{2t}\sum_{\s\in\zz} \Ebb_\P^\s\lkkk
\one_A(\gr(X_ {-t}), Q_{[-t,t]}  \gr(X_ t))_\hhh \rkkk ,\quad A\in {\ms G}_{[-t,t]}.
\en
Note that
$\mu(\ms X)= (\gr, e^{-2t (H-E)}\gr)=1$.
\bl{existence2}
There exists a unique probability measure $\mu_\infty$ on $(\ms X, \ov{\cal G})$ such that
$\mu_\infty\lceil_{\cal G}=\mu$. In particular, $\mu_\infty(A)=\mu(A)$, for every  $A\in {\ms G}_{[-t,t]}$ and $t\in\RR$.
\el
\proof
Let $\cup_{j=1}^\infty  A_j\in {\cal G}$ and  $A_i\cap A_j=\emptyset$ for $i \neq j$. Then there exists $t>0$
such that $\cup_{j=1}^\infty  A_j\in {\ms G}_{[-t,t]}$ by the definition of $\cal G$. Thus by the definition
of $\mu$ we have
\begin{align*}
\mu (\cup_{j=1}^\infty  A_j) = e^{2Et}e^{2t}\sum_{\s\in\zz}\Ebb_\P^\s[\one_{\cup_{j=1}^\infty A_j}
(\gr(X_ {-t}), Q_{[-t,t]}  \gr(X_ t))] = \sum_{j=1}^\infty \mu (A_j)
\end{align*}
by the Lebesgue dominated convergence theorem. Hence the set function $\mu$ on $(\ms X, {\cal G})$ is a
completely additive measure. Then the Hopf extension theorem implies that there exists a unique probability
measure $\mu_\infty$ on $(\ms X, \ov{\cal G})$ such that $\mu_\infty\lceil_{\cal G}=\mu$.
\qed
In order to show that $\mu_T(A)\to \mu_\infty(A)$ for every $A\in {\ms G}_{[-t,t]}$, we define the probability
measure $\rho_T$ on $(\ms X, {\ms G}_{[-T,T]})$ for $A\in \ms G_{[-t,t]}$ with $t\leq T$ by
\eq{rho}
 \rho_T(A)= e^{2Et}e^{2t}\sum_{\s\in\zz}\Ebb_\P^\s\lkkk \one_A
\lk \frac{\Phi_{T-t}(X_ {-t})}{\|\Phi_T\|}, Q_{[-t,t]} \frac{\Phi_{T-t}(X_ {t})}{\|\Phi_T\|}\rk\rkkk.
\en
The  family of probability measures $\rho_T^\La$ on $(\zz^\La, \ms B^\La)$ indexed by $\Lambda=\{t_0,...,t_n\}
\subset  [-T,T]$ is defined by
\eq{finiterho}
\rho_T^\La (A_0\times\cdots\times A_n) = e^{2Et}e^{2t}\sum_{\s\in\zz}\Ebb_\P^\s\lkkk
\lk
\prod_{j=0}^n
\one_{A_{j}}(X_ {t_j})
\rk\lk \frac{\Phi_{T-t}(X_ {-t})}{\|\Phi_T\|}, Q_{[-t,t]} \frac{\Phi_{T-t}(X_ {t})}{\|\Phi_T\|}\rk\rkkk
\en
for arbitrary $t$ such that $-T\leq -t\leq \ldots \leq t_0 \leq \ldots\leq t_n\leq t\leq T$. To show that
$\mu_T=\rho_T$, we prove that their finite dimensional distributions  coincide.
\bl{finitedistribution}
Let $\La=\{t_0,t_1,...,t_n\}$ and $A_0\times\cdots\times A_n\in \ms B^\Lambda$. Then it follows that
$$
\mu_T^\La(A_0\times\cdots\times A_n)=\rho_T^\La (A_0\times\cdots\times A_n).
$$
In particular, $\rho_T^\La$ is independent of the choice of $t$.
\el
\proof
By Corollary \ref{greenfunction} we see that
\begin{align*}
\mu_T^\La (A_0\times\cdots\times A_n)= \frac{1}{\|\Phi_T\|^2}
(\onee,e^{-(t_0+T){\PF}}\one_{A_0}e^{-(t_1-t_0){\PF}}\one_{A_1}\cdots \one_{A_n}e^{-(T-t_n){\PF}}\onee).
\end{align*}
Hence we have by the definition of $\Phi_{T-t}$ that
$$
\mu_T^\La (A_0\times\cdots\times A_n)= \frac{e^{2Et}}{\|\Phi_T\|^2} (\Phi_{T-t},e^{-(t_0+t){\PF}}
\one_{A_0}e^{-(t_1-t_0){\PF}}\one_{A_1}\cdots \one_{A_n}e^{-(t-t_n){\PF}} \Phi_{T-t}).
$$
By Corollary \ref{greenfunction} we have furthermore
\begin{align*}
\mu_T^\La (A_0\times\cdots\times A_n)
& =
{e^{2Et}e^{2t}} \sum_{\s\in\zz}\Ebb_{\P}^\s\lkkk \prod_{j=0}^n \one_{A_j}(X_ {t_j})\lk
\frac{\Phi_{T-t}}{\|\Phi_T\|} (X_ {-t}), Q_{[-t,t]} \frac{\Phi_{T-t}(X_ {t})}{{\|\Phi_T\|}}\rk  \rkkk\\
&=
\rho_T^\La(A_0\times\cdots\times A_n).
\end{align*}
Thus the lemma follows.
\qed

Denote $\zz^{(-\infty,\infty)}=\{\omega: (-\infty,\infty) \to \zz\}$.
\bl{identity}
Let $t\leq T$ and  $A\in \ms G_{[-t,t]}$. Then $\mu_T(A)=\rho_T(A)$.
\el
\proof
It is straightforward to see that the family of probability measures
$\mu_T^\La$, $\Lambda\subset \RR$,
on $(\zz^\La,\ms B ^\La)$ with $\#\La<\infty$ satisfies the Kolmogorov consistency condition:
$$
\mu_T^{\{t_0,...,t_n,s_1,...,s_m\}}(A_0\times\cdots\times A_n\times \prod^m \zz)
= \mu_T^{\{t_0,...,t_n\}}(A_0\times\cdots\times A_n).
$$
Let $\pi_\Lambda: \zz^{(-\infty,\infty)} \to \zz^\Lambda$ be the projection defined by $\pi_\Lambda(\omega)=
(\omega(t_0),\ldots,\omega(t_n))$ for $\omega\in \zz^{(-\infty,\infty)}$ and $\Lambda=\{t_0,\ldots,t_n\}$.
Then ${\ms A}_T= \{\pi^{-1}_\Lambda(E)\,|\,\Lambda\subset[-T,T],\#\Lambda<\infty,E\in \ms B^\Lambda\}$
is a finitely additive family of sets. Thus by the Kolmogorov extension theorem there exists a
unique probability measure $\mu_T^{(-\infty,\infty)}$ on $(\zz^{(-\infty,\infty)}, \s({\ms A}_T))$ such that
\eq{uni}
\mu_T^{(-\infty,\infty)}(\pi_\Lambda^{-1}(A_0\times\cdots\times A_n))=\mu_T^\La(A_0\times\cdots\times A_n)
\en
for all $\La\subset [-T,T]$ with $\#\La<\infty$ and $A_j\in \ms B$.
Note that $\zz^{(-\infty,\infty)}={\ms X}$ and
$\s({\ms A}_T)={\ms G}_{[-T,T]}$ follow.
On the other hand, we have
$$
\mu_T(\pi_\La^{-1}(A_0\times\cdots\times A_n))=\mu_T^\La(A_0\times\cdots\times A_n),
$$
and $\mu_T\lceil_{{\ms G}_{[-T,T]}}$ is a probability measure on $({\ms X}, {{\ms G}_{[-T,T]}})$.
Thus the uniqueness of
$\mu_T^{(-\infty,\infty)} $ satisfying \kak{uni} implies that
$\mu_T^{(-\infty,\infty)} =\mu_T\lceil_{{\ms G}_{[-T,T]}}$.
Observing that
$$
 \mu_T^\La(A_0\times\cdots\times A_n) = \rho_T^\La(A_0\times\cdots\times A_n) = \rho_T(\pi_\La^{-1}(A_0\times\cdots\times A_n))
$$
by Lemma \ref{finitedistribution}, we also see that
$\rho_T(A)=\mu_T^{(-\infty,\infty)}(A)$ for $A\in {\ms G}_{[-t,t]}$ by the uniqueness of $\mu_T^{(-\infty,\infty)}$
satisfying \kak{uni}.
Thus  together with
$\mu_T^{(-\infty,\infty)} =\mu_T\lceil_{{\ms G}_{[-T,T]}}$
we conclude that
$\mu_T(A)=\rho_T(A)$ for $A\in \ms G_{[-t,t]}$ and $t\leq T$.
\qed

\bt{gibbs}
Suppose $\hat h/\omega\in\LR$. Then the probability measure $\mu_T$ on $(\ms X, \ov{\cal G})$ converges  in local weak sense to $\mu_\infty$ as $T\to\infty$.
\et
\proof
By Lemma \ref{identity} it suffices to show that $\d \lim_{T\to\infty}\rho_T(A)=\mu_\infty(A)$ for every $A\in\ms G_{[-T,T]}$.
Since $\Phi_{T-t}/\|\Phi_T\|\to \gr$ strongly in $\LZ\otimes L^2(Q)$ as $T\to\infty$, we have
$\Phi_{T-t}(\s)/\|\Phi_T\|\to \gr(\s)$ for every $\s\in\zz$, strongly in $L^2(Q)$. Since $Q_{[-t,t]}$
is a bounded operator, it is seen that
\begin{align*}
\lim_{T\to \infty}\rho_T(A)&=\lim_{T\to \infty}e^{2t}e^{2Et}\sum_{\s\in\zz}
\Ebb_\P^\s \lkkk \lk \frac{\Phi_{T-t}(X_ {-t})}{\|\Phi_T\|}, Q_{[-t,t]}
\frac{\Phi_{T-t}(X_ t)}{\|\Phi_T\|}\rk \one _A\rkkk\\
&=
e^{2t}e^{2Et}\sum_{\s\in\zz}\Ebb_\P^\s \lkkk \lk \gr(X_ {-t}), Q_{[-t,t]} \gr(X_ t) \rk \one _A\rkkk =
\mu_\infty(A).
\end{align*}
Thus the theorem follows.
\qed

\subsection{The case of arbitrary $\eps>0$}
In the case when $\eps\not=1$ a parallel discussion to the previous section can be made. Since
$$
tH=\eps t \lk -\s_x\otimes\one+\one\otimes\frac{1}{\eps}\hf+\frac{\alpha}{\eps}\s_z\otimes\phi( h)\rk,
$$
by replacing $t$, $h$ and $\omega$ with $\eps t$, $h/\eps$ and $\omega/\eps$, respectively we have
\begin{eqnarray}
(\Phi, e^{-t\PF}\Psi)_\hhh = e^{\eps t} \sum_{\s\in\zz} \Ebb_\P^\s
\EEE \left[\ov{J_0^\eps \Phi(X_ 0)} e^{-(\alpha/\eps) \Phi_{\rm E}\lk \int_0^{\eps t}
X_ s j_s^\eps  h ds\rk }  J_t^\eps \Psi(X_ {\eps t}) \right].
\end{eqnarray}
Here $J_t^\eps$ and $j_t^\eps$ are defined by $\omega$ replaced by $\omega/\eps$.
Thus ${j_s^\eps}^\ast
j_t^\eps=e^{-|t-s|\omega/\eps}$ and ${J_s^\eps}^\ast J_t^\eps=e^{-|t-s|\hf/\eps}$.
Define the probability
measure $\mu_T^\eps$ on $(\ms X, \ov{\cal G})$ by
\eq{measureeps}
\mu_T^\eps (A)=\frac{e^{2\eps T}}{Z_{\eps T}} \sum_{\s\in\zz} \Ebb_\P^\s\lkkk \one_A
e^{\frac{\alpha^2}{2}\int_{-T}^{T} dt\int_{-T}^{ T} ds W (X_{\eps t},X_{\eps s}, t-s) }\rkkk,
\quad A\in \ov{\cal G}.
\en
Define also an additive set function on $(\ms X, {\cal G})$ by
$$
\mu^\eps(A)=e^{2E\eps t}e^{2\eps t}\sum_{\s\in\zz}
\Ebb_{\cal W}^\s[\one_A(\gr(X_{-\eps t}), Q_{[-\eps t,\eps t]}^{(\eps)} \gr(X_{\eps t})_\hhh],
\quad A\in \ms G_{[-\eps t,\eps t]},
$$
where $Q_{[-\eps t,\eps t]}^{(\eps)}=J_{-\eps t}^{\eps\ast} e^{\Phi_{\rm E}(-(\alpha/\eps)
\int_{-\eps t}^{\eps t}X_s j_s^\eps h ds)} J_{\eps t}^\eps $. In the same way as Lemma \ref{existence2} we
see that there exists a unique  probability measure $\mu_\infty^\eps$
on $(\ms X, \ov{\cal G})$ such
that $\mu_\infty^\eps \lceil_{\cal G}=\mu^\eps$.
Furthermore, it can be derived in a similar manner to Theorem
\ref{gibbs} that
\eq{eps}
\d \lim_{T\to\infty}\mu_T^\eps(A)=\mu_\infty^\eps (A),\quad A\in \ms G_{[-t,t]}.
\en
We summarize this in the theorem below.
\bt{gibbs2}
Suppose $\hat h/\omega\in\LR$.
Then the probability measure $\mu_T^\eps$ on $(\ms X, \ov{\cal G})$ converges  in local weak sense to $\mu_\infty^\eps$ as $T\to\infty$.
\et
We also write $\ggg$ for $\mu_\infty^\eps$ for notational convenience.

\subsection{Gibbs measure}
In this subsection we show that $\ggg  $ is a Gibbs measure on
$(\ms X,\ov{\cal G})$.
First we give some definitions and basic facts on Gibbs measures needed for this proof.

Let $(\Omega, {\FFF}, Q)$ be a probability space, and $\pro Y$ be a Markov process with c\`adl\`ag
paths on it. We write $\FFF_T=\s(Y_r, r\in [-T,T])$ and $\ms T_T=\s(Y_r, r\in [-T,T]^c)$. Let
${\ms V}:\BR\to \RR$ and ${\ms W}:\BR\times\BR\times\RR\to\RR$ be Borel measurable functions,
called external potential and pair potential, respectively. We call ${\ms V}$ an admissible
external potential whenever
\begin{align}
0<\Ebb_Q[e^{-\int_I{\ms V}(Y_s)ds}]<\infty
\end{align}
for every bounded interval $I\subset \RR$. Furthermore, we say that ${\ms W}$ is an admissible
pair interaction potential whenever
\begin{align}
\int_{\RR} \sup_{x,y\in\BR}|{\ms W}(x,y,s)|ds<\infty.
\end{align}

For the  admissible potentials $\ms V, \ms W$ and $0 < S \leq T$ define the functionals
\begin{align}
{\ms E}_T&=\int_{-T}^ T{\ms V}(Y_t) dt+
\left(\int_\RR ds\int_{-T}^T dt + \int_{-T}^T ds\int_\RR dt\right)
{\ms W}(Y_t,Y_s,|t-s|),\\
{\ms E}_{S,T}&=\int_{-T}^ T{\ms V}(Y_t) dt+
\left(\int_{-S}^Sds\int_{-T}^T dt + \int_{-T}^Tds\int_{-S}^S dt\right){\ms W}(Y_t,Y_s,|t-s|).
\end{align}
Also, define $Q_T^Y$ on $(\Omega,\FFF)$ for every $Y\in \Omega$ as the unique probability measure
such that $\Ebb_{Q_T^Y}[fg]=\Ebb_Q[f|\ms T_T](Y) g(Y)$, for every bounded $\FFF_T$-measurable
function $f$ and every bounded $\ms T_T$-measurable function $g$,
i.e.,
\eq{qya}
Q_T^Y[A]=\Ebb_{Q}[\one_A|\ms T_T](Y).
\en
\begin{definition}
{\rm
Suppose that ${\ms V}$ and ${\ms W}$ are admissible potentials.
\begin{enumerate}
\item[(1)]
A probability measure $P_T$ on $(\Omega, \FFF)$ is called a \emph{finite volume Gibbs measure}
for the interval $[-T,T]$ with respect to the reference measure $Q$ and the potentials ${\ms V}$
and ${\ms W}$ whenever for all $0<S<T$
\begin{enumerate}
\item[(i)]
$P_T\lceil_{\FFF_T}\ll Q\lceil_{\FFF_T}$
\item[(ii)]
for every bounded $\FFF$-measurable function $f$
\eq{gibbsdeffinite}
\Ebb_{P_T}[f|{\ms T}_S](Y)=\frac{\Ebb_{Q_S^Y}[fe^{-{\ms E}_{S,T}}]}{\Ebb_{Q_S^Y}[e^{-{\ms E}_{S,T}}]},
\quad \mbox{$P_T$-a.s.}
\en
\end{enumerate}

\item[(2)]
A probability measure $P$ is called a \emph{Gibbs measure} with respect to the reference measure
$Q$ and the potentials ${\ms V}$ and  ${\ms W}$ whenever for all $T>0$
\begin{enumerate}
\item[(i)]
$P\lceil_{{\FFF}_T}\ll Q\lceil_{\FFF_T}$
\item[(ii)]
for every bounded $\FFF$-measurable function $f$
\eq{gibbsdef}
\Ebb_P[f|{\ms T}_T](Y)=\frac{\Ebb_{Q_T^Y}[fe^{-{\ms E}_T}]}{\Ebb_{Q_t^Y}[e^{-{\ms E}_T}]},
\quad \mbox{$P$-a.s.}
\en
\end{enumerate}
\end{enumerate}
}
\end{definition}

A sufficient condition for $P_T$ to be a finite volume Gibbs measure and $P$ a Gibbs measure is as
follows.
\bp{gibbscriteria}
Let ${\ms V}$ and ${\ms W}$ be admissible potentials.
\begin{enumerate}
\item[(1)]
For every $T>0$
\eq{pt11}
\d dP_T=\frac{1}{Z_T} e^{-{\ms E}_{T,T}} dQ
\en
is a finite volume Gibbs measure for $[-T,T]$, where $Z_T$ denotes the normalizing constant.
\item[(2)]
Suppose that there exists a probability measure $P_\infty$ such that $P_t(A)\to P_\infty(A)$ as
$t\to\infty$ for all $A\in \FFF_T$, and $P_\infty\lceil_{\FFF_T}\ll Q
\lceil_{\FFF_T}$ for every $T$. Then $P_\infty$ is a Gibbs measure for the given potentials and
reference measure.
\end{enumerate}
\ep
\proof
For (1) see Proposition 4.1, for (2) Proposition 4.2 in \cite{lhb11}.
\qed

Consider on $\zz$ the Bernoulli measure
$$
\nu(\sigma) = \frac{1}{2} (\delta_{-1}(\sigma) + \delta_{+1}(\sigma)),\quad\s\in\zz,
$$
and define the probability measure $\mu_0$ on $(\ms X,\ov{\cal G})$  by
$$
{\mu_0}(A) = \Ebb_\nu\Ebb_{\cal W}^\s[\one_A], \quad A \in \ov{\cal G}.
$$
\bt{gibbsmeasure}
Suppose that $\hat h/\omega\in\LR$. Then the probability measure
$\ggg $ is a Gibbs measure on $(\ms X, \ov{\cal G})$ with respect to reference measure
$\mu_0$, zero external potential and pair interaction potential
$$W(X_{\eps t}, X_{\eps s}, |t-s|)=
\half X_{\eps t}X_{\eps s}\int_\BR e^{-|t-s|\omega(k)} |\hat h(k)|^2dk.$$
\et
\proof
The probability measure $\mu_T^\eps$ is a finite volume Gibbs measure by part (1) of Proposition
\ref{gibbscriteria} and \kak{measureeps}. By Theorem  \ref{gibbs2} we have that $\mu_T^\eps(A)\to
\ggg (A)$ as $T\to\infty$ for every $A\in \ms G_{[-t,t]}$, $t \leq T$, and
\begin{align*}
\ggg (A)&=e^{2E\eps t}e^{2\eps t}
\sum_{\s\in\zz}\Ebb_\P^\s\lkkk \lk \gr(X_ {-\eps t}), Q_{[-\eps t,\eps t]}^{(\eps)}
\gr(X_ {\eps t})\rk \one_A \rkkk\\
&\leq
2e^{2E\eps t}e^{2\eps t} \|Q_{[-\eps t,\eps t]}^{(\eps)}\|_{L^1(Q)} \mu_0(A)
\leq
2e^{2E\eps t}e^{2\eps t}e^{\alpha^2 t^2\|h\|^2} \mu_0(A).
\end{align*}
This bound is derived from \kak{44}. Hence $\ggg \lceil_{\FFF_t}\ll
\mu_0\lceil_{\FFF_t}$ follows for every $t>0$. Then the theorem follows by part (2) of Proposition
\ref{gibbscriteria}.
\qed

\section{Ground state properties}
\subsection{Expectations of functions of the form $\xi(\s)F(\phi(f))$}
In this section we use the Gibbs measure obtained above to derive ground state properties of the form \kak{ex2}
mentioned in  Section \ref{s11}. We start by considering ground state expectations of the form  $(\gr, \xi(\s) F(\phi(f))\gr)$ with suitable functions $F$ and $\xi$ expressed through expectations with respect to the path measure $\ggg$.
 By the parity symmetry
we know that
\eq{paritysymmetry}
(\gr,\s\gr)_{L^2(\zz;L^2(Q))}=(\grsb, \s_x\grsb)_{\CC^2\otimes\fff}=0.
\en
\subsubsection{Expectation of $\xi(\s)$}
\bt{grounstateexpectation}
Let  $f$ be  a ${\ms G}_{[-\eps t,\eps t]}$-measurable function on $\ms X$. Then
\eq{fkf3}
\EEEE [f] =e^{2E\eps t}e^{2\eps t}
\sum_{\s\in\zz}\Ebb_\P^\s\lkkk \lk \gr(X_ {-\eps t}), Q_{[-\eps t,\eps t]}^{(\eps)}\gr(X_ {\eps t})\rk f \rkkk.
\en
\et
\proof
Since for $A\in \ms G_{[-\eps t,\eps t]}$ we have
$$
\ggg (A)=e^{2\eps t}e^{2E\eps t}\sum_{\s\in\zz}
\Ebb_\P^\s \lkkk \lk \gr(X_ {-\eps t}), Q_{[-\eps t,\eps t]}^{(\eps)} \gr(X_ {\eps t}) \rk
\one _A\rkkk,
$$
\kak{fkf3} follows.
\qed
An immediate consequence of Theorem  \ref{grounstateexpectation} is the following.
\bc
{expectation}
Let  $f_j:\zz\to \CC$, $j=0,...,n$,  be bounded functions.
Then
\eq{finitedistribution2}
\EEEE \lkkk \prod_{j=0}^n f_j(X_ {\eps t_j}) \rkkk  =
(\gr, f_0e^{-(t_1-t_0)(H-E)}f_1\cdots e^{-(t_n-t_{n-1})(H-E)}f_n\gr).
\en
In particular, we have for all bounded functions $\xi, f$ and $g$ that
\begin{align}
\label{finitedistribution22}
\EEEE \lkkk \xi(X_0) \rkkk & =
(\gr,  \xi(\s)\gr),\\
\EEEE \lkkk f(X_t)g(X_s) \rkkk & =
(f(\s)\gr,e^{-|t-s|(H-E)}g(\s) \gr).
\end{align}
\ec
\proof
For $A_j\in\ms B$, $j=0,1,...,n$, it follows that
\begin{align*}
\EEEE \lkkk \prod_{j=0}^n \one_{A_j}(X_{\eps t_j})\rkkk
&=
e^{2\eps t}e^{2E\eps t}\sum_{\s\in\zz}\Ebb_\P^\s \lkkk \lk \gr(X_ {-\eps t}),
Q_{[-\eps t,\eps t]}^{(\eps)}  \gr(X_ {\eps t}) \rk
\prod_{j=0}^n \one_{A_j}(X_{\eps t_j}) \rkkk\\
&=
(\gr, \one_{A_0} e^{-(t_1-t_0)(H-E)}\one_{A_1}\cdots e^{-(t_n-t_{n-1})(H-E)}\one_{A_n}\gr).
\end{align*}
Hence \kak{finitedistribution2} is obtained.
\qed

\subsubsection{Expectation of $\xi(\s)F(\phi(f))$}
\bl{expectationoffield}
Let $F$ be a real-valued bounded function on $\RR$,  $f\in\LR$ and $\xi:\zz\to\CC$ be a bounded function.  Then
\begin{eqnarray*}
\lefteqn{
(e^{-TH}\onee,  \xi(\s) F(\phi(f)) e^{-TH}\onee) \non} \\
&&
= e^{2\eps T}\sum_{\s\in\zz}
\Ebb_\P^\s\Ebb_{\mu_{\rm E}}
\lkkk \xi(X_0) e^{-\frac{\alpha}{\eps}\Phi_{\rm E}(\int_{-\eps T}^{\eps T} X_s j_s h ds)} F(\Phi_{\rm E}(j_0 f))\rkkk.
\end{eqnarray*}
\el
\proof
By \kak{rep2} we have
$$
(e^{-TH}\xi(\s)F(\phi(f)) e^{-TH}\one_\hhh )(\s)=
e^{2\eps T}\Ebb_\P^\s \lkkk Q_{[-\eps T, 0]}^{(\eps)}
\xi(X_0) F(\phi(f))\Ebb_\P^{X_0}\lkkk Q_{[0,\eps T]}^{(\eps)}  \onee(X_{\eps T})\rkkk\rkkk.
$$
Here $Q_{[S,T]}^{(\eps)}={J_S^{\eps \ast} e^{\Phi_{\rm E}(-\frac{\alpha}{\eps}\int_S^T X_ s j_s^\eps h ds)}J_T^\eps}$.
Then in a similar manner to the proof of Corollary \ref{greenfunction} the lemma follows from the Markov property
of $\pro N$.
\qed
\bt{expectationoffield2}
Let $\hat h/\omega\in\LR$, $f\in\LR$ be real-valued,  $\xi:\zz\to\CC$  be a bounded function, and $\beta \in\RR$. Then
\begin{align}
(\gr, \xi(\s)e^{i\beta \phi(f)}\gr)=e^{-\frac{\beta ^2}{4}\|f\|^2}
\EEEE \lkkk \xi(X_0) e^{i\beta  \KI }\rkkk,
\end{align}
where $\KI $ is a random variable on $(\ms X,\ov{\cal G})$ given by
\eq{K}
\KI =-\frac{\alpha}{2}\int_{-\infty}^\infty (e^{-|r|\omega}\hat h,\hat f)  X_{\eps r} dr.
\en
\et
\proof
Note that
$$\d (\gr, \xi(\s) e^{i\beta\phi(f)}\gr)=
\lim_{T\to\infty}{(\grt, \xi(\s) e^{i\beta \phi(f)}\grt)}$$ and
by Lemma~\ref{expectationoffield} we see that
\begin{align*}
{\lk \grt, \xi(\s) e^{i\beta \phi(f)}\grt\rk}
= \frac{1}{Z_{\eps T}} e^{2\eps T}\sum_{\s\in\zz} \Ebb_\P^\s\Ebb_{\mu_{\rm E}}\lkkk
\xi(X_0) e^{-\frac{\alpha}{\eps}\Phi_{\rm E}(\int_{-\eps T}^{\eps T} X_s j_s h ds)} e^{i\beta  \Phi_{\rm E}(j_0 f)}
\rkkk.
\end{align*}
The expectation with respect to $\mu_{\rm E}$ can be computed explicitly and thus
\begin{align*}
&(\gr, \xi(\s)e^{i\beta\phi(f)} \gr)\\
&=
\lim_{T\to\infty} e^{-\frac{\beta^2}{4}\|f\|^2} \frac{1}{Z_{\eps T}}e^{2T}\sum_{\s\in\zz}
\Ebb_\P^\s \lkkk \xi(X_0) e^{\frac{\alpha^2}{2} \int_{- T}^{ T} dt \int_{-T}^{ T} W(X_{\eps t},X_{\eps s}, t-s) ds}
e^{-i\frac{\beta\alpha}{2}  \int_{- T}^{T}  (e^{-|s|\omega } \hat h, \hat f) X_{\eps s}ds} \rkkk\\
&=
\lim_{T\to\infty} e^{-\frac{\beta^2}{4}\|f\|^2} \Ebb_{\mu_T^\eps}
\lkkk \xi(X_0) e^{-i\frac{\beta\alpha}{2}  \int_{- T}^{ T}
 (e^{-|s|\omega}\hat  h, \hat f)X_{\eps s}ds }
\rkkk.
\end{align*}
Notice that $|\int_{-\infty}^\infty  X_{\eps s} (e^{-|s|\omega} \hat h, \hat f)ds|\leq
2 \|\hat h/\omega\|\|\hat f \|<\infty$. By the local weak convergence of $\mu_T$ and a similar telescoping
as in the proof of Theorem  \ref{main3} below, we obtain the desired result.
\qed
By using Theorem \ref{expectationoffield2} the functionals $(\gr, \xi(\s)F(\phi(f))\gr)$ can be represented in terms of averages with respect to the path measure $\ggg $. Consider the case when $F$ is a polynomial or  a Schwartz test function. We will show  in Corollary \ref{1} below that $\gr\in D(e^{+\beta N})$ for all $\beta>0$, thus  $\gr\in D(\phi(f)^n) $ for every $n\in{\Bbb N}$.

\bc{expectationoffield5}
Let $\hat h/\omega\in\LR$, $f\in\LR$ be real-valued,  and $\xi:\zz\to\CC$  a bounded function.
Also, let  $h_n(x)=(-1)^n e^{x^2/2}\frac{d^n}{dx^n}e^{-x^2/2}$  be the Hermite polynomial of order $n$. Then
\begin{align}
(\gr, \xi(\s)\phi(f)^n  \gr) =
i^n\EEEE \lkkk \xi(X_0)
 h_n\lk \frac{-i\KI }{\|f\|2^{-\han}}\rk
\rkkk (\|f\|2^{-\han})^n, \quad n \in{\Bbb N}.
\label{phimoments}
\end{align}
\ec
\proof
We have
\eq{gene}
e^{-\beta^2\|f\|^2/4}e^{i\beta \KI } =
\sum_{n=0}^\infty h_n\lk \frac{-i\KI }{\|f\|2^{-\han}}\rk \frac{(-\beta\|f\|2^{-\han})^n}{n!}.
\en
Hence
\eq{gene2}
\left.
\frac{1}{i^n}\frac{d^n}{d\beta^n}
e^{-\beta^2\|f\|^2/4}e^{i\beta \KI }\right \lceil_{\beta=0}=
i^n h_n\lk \frac{-i\KI }{\|f\|2^{-\han}}\rk (\|f\|2^{-\han})^n
\en
follows.  By \kak{gene2}  and the computation
\begin{align*}
(\gr, \xi(\s)\phi(f)^n\gr)
=\left.
\frac{1}{i^n}\frac{d^n}{d\beta^n} e^{-\frac{\beta^2}{4}\|f\|^2}
\EEEE [\xi(X_0)e^{i\beta \KI }]\right \lceil_{\beta=0},
\end{align*}
we obtain (\ref{phimoments}).
\qed
In the next corollary we give the path integral representation of $(\gr, \xi(\s)F(\phi(f))\gr)$ for $F\in \ms S(\RR)$, where $\ms S(\RR)$ denotes the space of rapidly decreasing, infinitely many times differentiable functions on $\RR$.

\bc{expectationoffield3}
Let $\hat h/\omega\in\LR$, $f\in\LR$ be real-valued,  $F\in {\ms S}(\RR)$,  and
$\xi:\zz\to\CC$  a bounded function. Then
\begin{align}
(\gr, \xi(\s) F(\phi(f)) \gr) = \EEEE \lkkk \xi(X_0) G\lk \KI \rk \rkkk,
\end{align}
where
$G=\check F \ast \check  g$ and $g(\beta)=e^{-\beta^2\|f\|^2/4}$.
\ec
\proof
Since $F(\phi(f))=\frac{1}{\sqrt{2\pi}} \int_{-\infty}^\infty \check F(\beta) e^{i\beta\phi(f)}d\beta$, we have
\begin{align}
(\gr, \xi(\s)F(\phi(f)) \gr) = \frac{1}{\sqrt{2\pi}} \int_{-\infty}^\infty \check F(\beta)
e^{-\frac{\beta ^2}{4}\|f\|^2} \EEEE \lkkk \xi(X_0) e^{i\beta  \KI }\rkkk d\beta.
\end{align}
Thus the corollary follows.
\qed

\subsubsection{Field fluctuations in the ground state}
The field fluctuations in the ground state are defined for every real-valued function $f\in\LR$ by
\begin{align}
F_\alpha (f)=(\gr, \phi(f)^2\gr)-(\gr, \phi(f)\gr)^2.
\end{align}
More generally, we also consider fluctuations of the form
\eq{fl2}
G_\alpha(f)=(\gr, (\s\phi(f))^2\gr)-(\gr, \s\phi(f)\gr)^2.
\en

\bc{expectations3}
Let $\hat h/\omega\in\LR$ and $f\in\LR$ be a real-valued function.
Then
\begin{itemize}
\item[(1)]
$\d (\gr, \s\phi(f)\gr)=\EEEE [X_0 \KI ]$,
\item[(2)]
$\d (\gr, (\s \phi(f))^2 \gr)=\EEEE [(X_0\KI )^2] + \half\|f\|^2$.
\end{itemize}
In particular,
\begin{itemize}
\item[(3)]
$G_\alpha(f)=\Ebb_{\ggg}[(X_0\KI )^2] -\lk \Ebb_{\ggg}[X_0\KI ]\rk^2 +\half\|f\|^2$,
\item[(4)]
$\d (\gr, \phi(f)\gr)=0$ and $\d F_\alpha(f)=
(\gr, \phi(f)^2 \gr)=\EEEE [\KI ^2] + \half\|f\|^2$,
\item[(5)]
whenever $f\not\equiv 0$, we furthermore have that (i) $F_\alpha(f)>0$ and $F_\alpha(f)\geq F_0(f)$,
(ii) $G_\alpha(f)>0$ and $G_\alpha(f)\geq G_0(f)$.
\end{itemize}
\ec
\proof
Statements (1)-(3) easily follow from Corollary \ref{expectationoffield5}, which imply (4) for
$\sigma = 1$. Using Schwarz inequality, we obtain (5ii), while (5i) is clear by (4). \qed
Note that
to prove (1)-(2) of Corollary \ref{expectations3} we can proceed, alternatively, to derive first the equality $(\gr,\phi(f)\gr)=
\Ebb_{\ggg}[\KI ]$ by using Corollary \ref{expectationoffield5}, and from
$\EEEE [X_s] = (\gr, \s\gr)$ to further obtain that
\eq{do}
\d(\gr, \phi(f)\gr)=-\frac{\alpha}{2}(\hat h/\omega, \hat f)(\gr, \s\gr)=0.
\en
Notice that  $X_0^2=1$. Thus in Corollary \ref{expectations3} we have equivalently
$\EEEE [(X_0\KI )^2]=\EEEE [\KI ^2]$.

\subsection{Gaussian decay and exponential moments of the field operator}
\subsubsection{Gaussian decay of the field operator}
In this section we show that $(\gr, e^{\beta \phi(f)^2}\gr)<\infty$ for some $\beta>0$.
\bl{gaussian1}
Let $\hat h/\omega\in\LR$ and $f\in \LR$ be a real-valued function. Then for $\beta>0$
we have
\eq{gauss}
(\gr, e^{-\beta\phi(f)^2}\gr)=
\frac{1}{\sqrt{1+\beta\|f\|^2}} \, \ee{e^{-\frac{\beta K^2(f)}{1+\beta\|f\|^2}}}.
\en
\el
\proof
By Theorem \ref{expectationoffield2} we see that
\begin{align*}
(\gr, e^{-(k^2/2)\phi(f)^2}\gr)
&=
\frac{1}{\sqrt{2\pi}} \int_\RR e^{-\beta^2/2}(\gr, e^{i\beta k \phi(f)}\gr)dk \\
&=
\frac{1}{\sqrt{2\pi}} \int_\RR e^{-k^2/2}e^{-\beta^2 k^2 \|f\|^2/4}\ee{e^{i\beta k K(f)}}dk\\
&=
\frac{1}{\sqrt{1+\beta^2\|f\|^2/2}} \, \ee{e^{-\frac{\beta^2 K^2(f)/2}{1+\beta^2\|f\|^2/2}}}.
\end{align*}
Replacing $\beta^2/2$ by $\beta$ completes the proof of the lemma.
\qed
\bt{gaussian decay}
Let $\hat h/\omega\in\LR$ and $f\in \LR$ be a real-valued function. If
$-\infty<\beta<1/\|f\|^2$, then
$\gr\in D(e^{(\beta/2)\phi(f)^2})$ and
\eq{gauss3}
\|e^{(\beta/2)\phi(f)^2}\gr\|^2
=
\frac{1}{\sqrt{1-\beta\|f\|^2}}\, \ee{e^{\frac{\beta K^2(f)}{1-\beta\|f\|^2}}}.
\en
\et
\proof
The proof is a modification of \cite[Theorem 10.12]{hir04}.

Let $B=\{z\in\CC| |z|<1/\|f\|^2\}$, $\CC_+=\{z|\,\Re z > 0\}$  and $\CC_-=\{z|\,\Re z < 0\}$.
Consider
\eq{r0}
\rho(z)=
\frac{1}{\sqrt{1+z\|f\|^2}} \, \ee{e^{-\frac{z K^2(f)}{1+z\|f\|^2}}},
\en
for $z>0$. Then $\rho(z)$ can be analytically continued to $\CC_+\cup B$, since
$|\KI|\leq |\alpha|\|f\| \|\hat h/\omega\|$ uniformly in paths. We denote this extension by
$\bar\rho(z)$. Let $w\in \RR\cap B$ and consider the ball $B_\delta(w)=\{z\in\CC|\,|z-w|<\delta\}$.
Take any $\delta<1/\|f\|^2$ such that for $w$ we have $B_\delta(w)\cap \CC_- \cap B\not=\emptyset$.
We expand $\bar \rho(z)$ as
\eq{r1}
\bar\rho(z)=
\sum_{n=0}^\infty (z-w)^n b_n(w),\quad z\in B_\delta(w)\cap B
\en
and get in particular
\eq{ab1}
\bar\rho(z) = \sum_{n=0}^\infty |z-w|^n |b_n(w)|<\infty
\en
for $z\in B_\delta(w)\cap B$. On the other hand, $\CC_+\ni z\mapsto (\gr, e^{-z\phi(f)^2}\gr)\in\CC$
is differentiable on $\CC_+$, since $\gr\in D(\phi(f)^2)$, and is thus analytic on $\CC_+$. We have
\eq{r2}
(\gr, e^{-z\phi(f)^2}\gr)=
\sum_{n=0}^\infty (z-w)^n \frac{1}{n!}\int_0^\infty (-\lambda)^ne^{-w\lambda}dE_\lambda,\quad z\in\CC_+.
\en
Here  $E_\lambda$ denotes the spectral measure of $\half \phi(f)^2$ with respect to $\gr$. Comparing
\kak{r1},  \kak{r2} and $\bar \rho(z)=(\gr, e^{-z\phi(f)^2}\gr)$ for $z\in\CC_+$, we conclude that
\eq{r3}
b_n(w)=\frac{1}{n!}\int_0^\infty (-\lambda)^ne^{-w\lambda}dE_\lambda.
\en
Substituting \kak{r3} into \kak{r1} we have
\eq{r4}
\bar \rho(z)=
\sum_{n=0}^\infty (z-w)^n
\frac{1}{n!}\int_0^\infty (-\lambda)^n e^{-w\lambda}dE_\lambda,\quad z\in B_\delta(w)\cap B
\en
where the right hand side is absolutely convergent for every $z\in B_\delta(w)\cap B$. Thus by \kak{ab1} for
$z\in B_\delta(w)\cap B\cap \RR$ we have
\begin{align*}
\int_0^M\!\! \!\!e^{-z\lambda}dE_\lambda
\leq\sum_{n=0}^\infty \frac{|z-w|^n}{n!}\left|\int_0^M \!\!\!\!(-\lambda)^n e^{-w\lambda}dE_\lambda\right|
\leq
\sum_{n=0}^\infty \frac{|z-w|^n}{n!}\left|
\int_0^\infty \!\!\!\!
(-\lambda)^n e^{-w\lambda}dE_\lambda\right|
<\infty,
\end{align*}
which implies that $\lim_{M\to\infty} \int_0^Me^{-z\lambda}dE_\lambda<\infty$ for these $z$. The monotone
convergence theorem then gives $\int_0^\infty e^{-z\lambda} dE_\lambda<\infty$, hence $\gr\in D(e^{-(z/2)
\phi(f)^2})$  and
\eq{r6}
\| e^{-(z/2)\phi(f)^2}\gr\|^2=\bar\rho(z), \quad z\in B_\delta(w)\cap B\cap \RR.
\en
Since for every $\delta<1/\|f\|^2$ there exists $w\in\RR\cap B$ such that  $\CC_-\cap B\cap B_\delta(w)\neq
\emptyset$, the proof of the theorem is complete.
\qed
From Theorem \ref{gaussian decay} it is immediate to get the limit of  $\|e^{(\beta/2)\phi(f)^2}\gr\|$.
\bc{gaussian nodecay}
Suppose that $\hat h/\omega\in\LR$ and $f\in\LR$ is a real-valued function. Then
$$
\lim_{\beta\uparrow 1/\|f\|^2} \|e^{(\beta/2)\phi(f)^2}\gr\|=\infty.
$$
\ec

In the previous section we investigated the moments of $\phi(f)$ of positive integer order. By using Lemma
\ref{gaussian1} also the moments of fractional order can be derived. Define $|\phi(f)|^s=(\phi(f)^2)^{s/2}$
for $0\leq s\leq 2$, and let $\lambda$ be the L\'evy measure on $\RR\setminus \{0\}$ such that $\int_0^\infty
(1-e^{-yu}) \lambda (dy) = u^{s/2}$ for $u > 0$, i.e.,
$$
\lambda(dy)=\frac{s}{2\Gamma(1-s/2)} y^{-1-\frac{s}{2}}1_{(0,\infty)}(y)dy,
$$
corresponding to the $s/2$-subordinator. Let $\Lambda_\alpha=\| |\phi(f)|^{s/2}\gr\|^2$.
\bc{fp}
Suppose that $\hat h/\omega\in\LR$ and $f\in\LR$ is a real-valued function. Then for $0< s< 2$,
\eq{ab2}
\Lambda_\alpha=\ee{\int_0^\infty \lk 1-\frac{1}{\sqrt{1+\beta\|f\|^2}}
e^{-\frac{\beta \KI^2}{1+\beta\|f\|^2}}\rk \lambda(d\beta)}.
\en
In particular, $\Lambda_0\leq \Lambda_\alpha$ follows.
\ec
\proof
Notice that
\begin{eqnarray*}
\lk 1-\frac{1}{\sqrt{1+\beta\|f\|^2}} e^{-\frac{\beta \KI^2}{1+\beta\|f\|^2}}\rk
\beta^{-1-\frac{s}{2}}
&\leq& \lk 1-\frac{1}{\sqrt{1+\beta\|f\|^2}}
e^{-\frac{\beta \alpha^2  \|f\|^2  \|\hat h/\omega\|^2/4}{1+\beta\|f\|^2}}\rk
\beta^{-1-\frac{s}{2}} \\
&=&
\eta(\beta).
\end{eqnarray*}
In a neighborhood of $\beta=0$ it holds that $\eta(\beta)=\beta ^{-s/2}+ o(\beta)$
locally uniformly. Then $\eta(\beta)$ is integrable in this neighborhood, and since
$\eta(\beta)\leq {\rm const} \,\beta^{1+s/2}$, $\eta\in L^1([0,\infty))$ follows. Then \kak{ab2} is
immediate from Lemma \ref{gaussian1} by using the Fubini theorem, and the inequality follows from
$$
\Lambda_0 = \int_0^\infty \lk 1-\frac{1}{\sqrt{1+\beta\|f\|^2}}\rk \lambda(d\beta).
$$
\qed

\begin{remark}
\label{jozsef}
{\rm
A simple computation shows that if $U$ is a real-valued Gaussian random variable with mean $m$ and variance
$\sigma^2$, then whenever $\theta < 1/(2\sigma^2)$, we have
$$
\Ebb_G[e^{\theta U^2}] = \frac{1}{\sqrt{1-2\theta\sigma^2}} \, e^{\frac{m^2\theta}{1-2\theta\sigma^2}},
$$
where the expectation is taken with respect to this Gaussian measure. A comparison with \kak{gauss3} implies
that there exists thus a real-valued Gaussian random variable $U$ with mean $K(f)$ (or $-K(f)$) and variance
$\|f\|^2/2$ such that
$$
\|e^{\beta\phi(f)^2}\gr\|^2 = \Ebb_G[e^{\beta U^2}].
$$
}
\end{remark}
For some consequences see Section \ref{vH} below.

\subsubsection{Exponential moments of the field operator}
Theorem \ref{gaussian decay} says that $\|e^{(\beta/2)\phi(f)^2}\gr\|<\infty$. Using this fact we can
obtain explicit formulae of the exponential moments $(\gr, e^{\beta\phi(f)}\gr)$ of the field.
\bc{expmoment}
If $\hat h/\omega\in\LR$ and $f\in\LR$ is a real-valued function, then $\gr\in D(e^{\beta\phi(f)})$ and
\begin{align}
&
\label {dd}
(\gr,e^{\beta\phi(f)}  \gr) = (\gr,\cosh(\beta\phi(f))  \gr) =
e^{\frac{\beta^2}{4}\|f\|^2} \EEEE \lkkk  e^{\beta\KI }\rkkk,\\
&\label{ddd}
(\gr,\s e^{\beta\phi(f)}  \gr) = (\gr,\s \sinh(\beta\phi(f))  \gr) =
e^{\frac{\beta^2}{4}\|f\|^2} \EEEE \lkkk  X_0 e^{\beta\KI }\rkkk.
\end{align}
\ec
\proof
For simplicity we reset $\beta f$ to $f$.  By using the generating function $e^{xy-\frac{1}{2}y^2} =
\sum_{n=0}^\infty h_n(x)\frac{y^n}{n!}$  of the Hermite polynomials, summation in (\ref{phimoments})
gives
\begin{align}
\lim_{M\to\infty} (\gr, \sum_{n=0}^M  \frac{1}{n!}\phi(f)^n\gr)=
e^{\frac{1}{4}\|f\|^2} \EEEE \lkkk  e^{\KI }\rkkk.
\end{align}
We need to check that the left hand side converges to $(\gr, e^{\phi(f)}\gr)$. Notice that by the spin
flip property (2) in Lemma \ref{flip}, $(\gr, \phi(f)^n\gr)=0$ for odd $n$.  Hence it suffices to show
the convergence of $(\gr, \sum_{n=0}^M  \frac{1}{(2n)!}\phi(f)^{2n}\gr)$ as $M\to\infty$. By
Theorem~\ref{gaussian decay} we have that  $\|e^{\phi(f)^2/(4\|f\|^2)}\gr\|<\infty$. Let $E_\lambda$ be
the spectral measure of $\phi(f)$ with respect to $\gr$. Then
\begin{align*}
\sum_{n=0}^M  \frac{1}{(2n)!}(\gr, \phi(f)^{2n}\gr) =
\int_\RR \sum_{n=0}^M  \frac{1}{(2n)!}\lambda ^{2n} e^{-\lambda ^2/(4\|f\|^2)}
e^{\lambda^2/(4\|f\|^2)}dE_\lambda.
\end{align*}
Since $e^{\lambda^2/(4\|f\|^2)}$ is integrable by Theorem \ref{gaussian decay},
$\sum_{n=0}^M  \frac{1}{(2n)!}\lambda ^{2n} e^{-\lambda ^2/(4\|f\|^2)}$ is monotonously increasing to
$\cosh (\lambda )e^{-\lambda ^2/(4\|f\|^2)}$ as $M\uparrow\infty$, which is a bounded function, hence the
monotone  convergence theorem yields that $\lim_{M\to\infty}\int_\RR \sum_{n=0}^M  \frac{1}{n!}\lambda ^n
dE_\lambda = \int_\RR e^{\lambda}dE_\lambda<\infty$, which implies $\gr\in D(e^{\phi(f)})$ and \kak{dd}.
Equality \kak{ddd} is derived in a similar way.
\qed

\subsection{Van Hove representation}
\label{vH}
In Remark \ref{jozsef} we pointed out that the expectation of the field operator $\phi(f)$ in the ground
state $\gr$ can be realized as an expectation of a  Gaussian random variable. Here we show that this allows
another representation of the ground state.

The \emph{van Hove Hamiltonian} is defined by the self-adjoint operator
\eq{vh1}
\vh(\hat g)=\hf +\phib(\hat g)
\en
in Fock space $\fff$. Suppose that $\hat g /\omega\in \LR$ and define the conjugate momentum by
$$\pi_{\rm b}(\hat g)=\frac{i}{\sqrt{2}} \int \lk \add(k) \frac{\hat g(k)}{\omega(k)}-
a(k) \frac{\hat g(-k)}{\omega(k)} \rk dk.$$
Then
\eq{vh2}
e^{i\pi_{\rm b}(\hat g)} \vh(\hat g) e^{-i\pi_{\rm b}(\hat g)}=\hf -\half \|\hat g/\sqrt \omega\|^2
\en
and the ground state of $\vh(\hat g)$ is given by
$$\grvh(\hat g)=e^{-i\pi_{\rm b}(\hat g)}\Omega_{\rm b}.$$
 On the other hand, clearly the
spin-boson Hamiltonian  $H$ with $\eps=0$ is the direct sum of van Hove Hamiltonians since
$$
H=\mmm {\hf+\alpha\phib(\hat h)} 0 0 {\hf-\alpha\phib(\hat h)}
$$
(see Section 2) and $\hf\pm \alpha\phib(\hat h)$ are equivalent.
Therefore the ground state of $H$ with $\eps=0$ can
be realized as $\gr=\vvv{\grvh(\alpha \hat h) \\ \grvh(-\alpha \hat h)}$.
 Thus in this case
\eq{vh6}
(\gr, e^{i\beta\phi(f)}\gr)_\hhh
=\half \sum_{x=\pm 1} 
(\grvh(x\alpha \hat h), e^{i\beta\phi_{\rm b}(\hat f)}\grvh(x\alpha \hat h))_\fff.
\en
and the right hand side above equals
\eq{vh7}
(\Omega_{\rm b}, e^{i\beta(\phib(\hat f)+\alpha(\hat h/\omega, \hat f))}\Omega_{\rm b})_\fff=
e^{-\beta^2\|\hat f\|^2/4+i\beta \alpha (\hat h/\omega, \hat f)}.
\en

When $\eps\not=0$ we can derive similar but non-trivial representations. Define the random boson
field operator
\eq{vh3}
\Psi(\hat f)=\phib(\hat f)+K(f)
\en
on $\fff$, where $K(f)$ is the random variable on $\ms X$ defined by \kak{K}. Then we see that
\begin{align}
&(\Omega_{\rm b}, \Psi(\hat f)\Omega_{\rm b})=K(f),\\
&
(\Omega_{\rm b}, \Psi(\hat f)^2\Omega_{\rm b})-(\Omega_{\rm b}, \Psi(\hat f)\Omega_{\rm b})^2=\|\hat f\|^2/2,\\
&(\Omega_{\rm b}, e^{i\beta \Psi(\hat f)}\Omega_{\rm b})
=e^{-\beta^2\|f\|^2/4+i\beta\KI}.
\end{align}
 Let
\eq{vh5}
\chi=-\frac{\alpha}{2} \omega(k) \hat h(k) \int_{-\infty}^\infty e^{-|s|\omega(k) }X_{\eps s} ds.
\en
Note that $\chi\in\LR$
\eq{kii}
\KI=(\chi,\hat f),
\en
moreover, $\chi/\omega\in\LR$, whenever $\hat h/\omega\in\LR$, and  $\chi=\alpha \s \hat h$ for $\eps=0$. We define
the \emph{random van Hove Hamiltonian} by $\vh(\chi)$.
\bt{rep11}
If $\hat h/\omega\in\LR$, then
\begin{align}
(\gr, e^{i\beta\phi(f)}\gr)=
\ee {(\Omega_{\rm b}, e^{i\beta\Psi(\hat f)}\Omega_{\rm b})}=
\ee {(\grvh(\chi), e^{i\beta\phib(\hat f)}\grvh(\chi))}.
\end{align}
\et
\proof
The first equality can be directly derived from Theorem \ref{expectationoffield2}. The second equality
follows from  $e^{i\pi_{\rm b}(\chi)}\Psi(\hat f) e^{-i\pi_{\rm b}(\chi)}=\phib(\hat f)$.
\qed
\bc{i}
Suppose $\hat h/\omega\in\LR$ and $F\in \ms S(\RR)$. Then we have
\begin{align}
&(\gr, F(\phi(f))\gr)=\ee {(\Omega_{\rm b}, F(\Psi(\hat f))\Omega_{\rm b})}=\ee {(\grvh(\chi), F(\phi(\hat f))\grvh(\chi))},\\
&
\|e^{\beta\phi(f)^2/2}\gr\|^2=\ee {\| e^{\beta\Psi(\hat f)^2/2}\Omega_{\rm b}\|^2}=
\ee {\|  e^{\beta\phib(\hat f)^2/2}\grvh(\chi)\|^2}.
\end{align}
\ec
\proof
This is proven from Corollary \ref{expectationoffield3} and Theorem \ref{gaussian decay}.
\qed

\subsection{Expectations of second quantized operators}
\subsubsection{General results}
In this section we consider  expectations of the form $(\gr, e^{-\beta d\Gamma(\rho(-i\nabla))}\gr)$,
where $\rho$ is a real-valued multiplication operator given by the function $\rho$. An important
example is $\rho=\one$ giving the boson number operator  $N=d\Gamma(\one)$.

In a similar way to \cite[Section 3.2]{ghps09} we obtain the expression
\eq{number}
\frac{(\Phi_T,  \xi(\s) e^{-\beta \nn} \Phi_T)}{\|\Phi_T\|^2} = \Ebb_{\mu_T^\eps}
\lkkk \xi(X_0)e^{-{\alpha^2}\int_{-T}^0 dt \int_0^{T} W^{\rho,\beta} (X_{\eps t}, X_{\eps s}, t-s)ds}\rkkk,
\en
where
$$
W^{\rho,\beta}(x,y,T)=\frac{xy}{2}\int_\BR |\hat h(k)|^2e^{-|T|\omega(k) }(1-e^{-\beta\rho(k)}) dk.
$$
Denote
\eq{denote}
W_\infty^{\rho,\beta} = \int_{-\infty}^0 dt \int_0^\infty W^{\rho,\beta} (X_{\eps t}, X_{\eps s}, t-s) ds.
\en
Notice that $|W_\infty^{\rho,\beta}| \leq \|\hat h/\omega\|^2/2<\infty$,
uniformly in the paths in $\ms X$.

\bt{main3}
Suppose that $\hat h/\omega\in \LR$ and $\xi:\zz\to\CC$ is a bounded function. Then
\eq{fkf4}
(\gr, \xi(\s) e^{-\beta \nn}\gr)= \EEEE \lkkk
\xi(X_0) e^{-\alpha^2{W_\infty^{\rho,\beta}}}\rkkk,\quad \beta>0.
\en
\et
\proof
This is shown by using Theorem \ref{gibbs} and telescoping.
For a shorthand we write
$W _T^{\rho,\beta}= \int_{-T}^0ds\int_0^{T} W^{\rho,\beta}(X_{\eps t}, X_{\eps s}, t-s) dt$. Note that for
every $\delta >0$ there is $S_\delta $ such that $|W_T^{\rho,\beta} -W_\infty^{\rho,\beta}| \leq \delta $
for all $T>S_\delta $, uniformly in the paths, and write
\begin{align*}
&\Ebb_{\mu_T^\eps}\lkkk \xi(X_0)e^{-\alpha^2{W_T^{\rho,\beta}}}\rkkk -
\EEEE \lkkk \xi(X_0)e^{-\alpha^2{W_\infty^{\rho,\beta}}}\rkkk\\
&=
\Ebb_{\mu_T^\eps}\lkkk \xi(X_0)e^{-\alpha^2{W_T^{\rho,\beta}}}\rkkk
- \Ebb_{\mu_T^\eps}\lkkk
\xi(X_0) e^{-\alpha^2{W_\infty^{\rho,\beta}}}\rkkk \\
&\hspace{1cm}+ \Ebb_{\mu_T^\eps}\lkkk \xi(X_0)e^{-\alpha^2{W_\infty^{\rho,\beta}}}\rkkk -
\EEEE \lkkk \xi(X_0)e^{-\alpha^2{W_\infty^{\rho,\beta}}}\rkkk.
\end{align*}
We have
\begin{equation}
\label{upper}
\left|\Ebb_{\mu_T^\eps}\lkkk \xi(X_0)e^{-\alpha^2{W_T^{\rho,\beta}}}\rkkk -
\Ebb_{\mu_T^\eps}\lkkk \xi(X_0)e^{-\alpha^2{W_\infty^{\rho,\beta}}}\rkkk\right|
\leq C\delta
\end{equation}
with a constant $C$. The second term can be evaluated as
\begin{eqnarray}
\lefteqn{
\left|\Ebb_{\mu_T^\eps}\lkkk \xi(X_0)e^{-\alpha^2{W_\infty^{\rho,\beta}}}\rkkk -
\EEEE \lkkk \xi(X_0)e^{-\alpha^2{W_\infty^{\rho,\beta}}}\rkkk\right|\non } \\
&&
\label{b1}
\leq
\left|\Ebb_{\mu_T^\eps}\lkkk \xi(X_0)e^{-\alpha^2{W_\infty^{\rho,\beta}}}\rkkk -
\Ebb_{\mu_T^\eps}\lkkk \xi(X_0)e^{-\alpha^2{W_{S_\delta }^{\rho,\beta}}}\rkkk \right|\\
&&\label{b2}
\qquad + \left|\Ebb_{\mu_T^\eps}\lkkk \xi(X_0)e^{-\alpha^2{W_{S_\delta }^{\rho,\beta}}}\rkkk -
\EEEE \lkkk \xi(X_0)e^{-\alpha^2{W_{S_\delta }^{\rho,\beta}}}\rkkk\right |\\
&&
\label{b3}
\qquad +\left| \EEEE \lkkk \xi(X_0)e^{-\alpha^2{W_{S_\delta }^{\rho,\beta}}}\rkkk -
\EEEE \lkkk \xi(X_0)e^{-\alpha^2{W_\infty^{\rho,\beta}}}\rkkk\right|.
\end{eqnarray}
For \kak{b1} and \kak{b3} we have again the same upper bound as in (\ref{upper}), and \kak{b2}
goes to zero as $T\to \infty$ by Theorem \ref{gibbs}.
\qed

\subsubsection{Super-exponential decay of the  boson number}
In this section we discuss the expectation of $e^{-\beta\n}$, which can be obtained by
a minor modification of Theorem \ref{main3}.
 \bc{nn}
Suppose that $\hat h/\omega\in\LR$ and $\xi:\zz\to\CC$ is a bounded function.
Then
\begin{align}
\label{nnnn}
(\gr, \xi(\s) e^{-\beta \n }\gr)=\EEEE \lkkk
\xi(X_0) e^{-\alpha^2(1-e^{-\beta})W_\infty} \rkkk,
\end{align}
where \eq{denote2}
W_\infty =\int_{-\infty}^0 dt \int_0^\infty W (X_{\eps t}, X_{\eps s}, t-s) ds.
\en
\ec
\proof
By replacing $\rho$ by $\one$ in Theorem \ref{main3}, the claim readily follows.
\qed

The following result says that the distribution of the number of bosons in the ground state has a
super-exponentially short tail.
\bc{1}
If $\hat h/\omega\in \LR$, then $\gr\in D(e^{\beta \n})$ for all $\beta \in\CC$
and
\eq{N}
(\gr, e^{\beta N}\gr)=\ee{e^{-\alpha^2(1-e^\beta)W_\infty}}
\en
follows. In particular, $\gr\in D(e^{+\beta N})$ for all $\beta>0$.
 \ec
\proof
The proof is similar to that of Theorem \ref{gaussian decay} and \cite[Theorem 10.12]{hir04}, and is
left to the reader.
\qed

\bc{-1}
Suppose $\hat h/\omega\in \LR$.
Then
\begin{align}
&\label{n1}(\gr, (-1)^N\gr)=\EEEE \lkkk e^{-2\alpha^2W_\infty}\rkkk,\\
&\label{n2}(\gr, \xi(\s) (-1)^N\gr)=
\EEEE \lkkk
\xi(X_0) e^{-2\alpha^2W_\infty}\rkkk.
\end{align}
In particular, it follows that
\begin{align}
&\label{n3}(\gr, (-1)^N\gr)=\EEEE \lkkk
e^{-2\alpha^2W_\infty}\rkkk\geq e^{-\alpha^2\|\hat h/\omega\|^2}>0,\\
&\label{n4}(\gr, \s(-1)^N\gr)=\EEEE \lkkk
X_0 e^{-2\alpha^2W_\infty}\rkkk=-1<0.
\end{align}
\ec
\proof
Equality \kak{n1} is  derived from \kak{nnnn} with $\xi(\s)=1$ and $\beta=-i\pi$, and \kak{n2} with $\xi(\s)=\s$. Equality \kak{n3} follows from the estimate of the right hand side of \kak{n2}. Noticing that $\grsb\in \hhh_{-}$, we obtain $P\grsb=\s_x(-1)^N\grsb=-\grsb$. In particular, this gives
$$
\EEEE \lkkk  X_0 e^{-2\alpha^2W_\infty}\rkkk=(\gr, \s(-1)^N\gr)=(\grsb, P\grsb)=-1.
$$
\qed

\subsubsection{Moments of the boson number operator}
We can derive the expectation of $\n^m$, $m=1,2,...$,  with respect to the ground state $\gr$  by using Corollary \ref{nn}.
\bc{m11}
Suppose that  $\hat h/\omega\in \LR$.
Then
\eq{m}
(\gr, \n^m\gr)=
\sum_{r=1}^m a_r(m)\alpha^{2r}
\EEEE \lkkk W_\infty^r\rkkk,\quad m=1,2,3,...
\en
where
$a_r(m)=\frac{(-1)^r}{r!}\sum_{s=1}^r(-1)^s  {r\choose  s}s^m$ are the Stirling numbers of the second kind.
\ec
\proof
It can be checked that
$$
\frac{d^m}{d\beta^m} e^{-C(1-e^{-\beta})}=
(-1)^m \sum_{r=1}^m a_r(m)e^{-r\beta}
(-C)^r e^{-a(1-e^{-\beta})}.
$$
Then the corollary follows from $(\gr, N^m\gr)=(-1)^m \frac{d^m}{d\beta^m}(\gr, e^{-\beta \n}\gr)\lceil_{\beta=0}$ and  Corollary \ref{nn}.
\qed

\subsection{A relation between the expectations of $\s\phi(f)$ and $N$}
By the results obtained in the previous subsections we can derive an inequality connecting the expectations $(\gr, \s\phi(f)\gr)$ and $(\gr, \n\gr)$.
\bc{yume3}
Suppose that $\hat h/\omega\in \LR$,  $f\in\LR$ is real-valued, and $\xi:\zz\to\CC$ a bounded function. Then
\begin{align}
(\gr, \xi(\s)\phi(f) \gr)
=
{\alpha}\int_\BR (\xi(\s)\gr, (H-E+\omega(k))^{-1}\s\gr)\ov{\hat h}(k) \hat f(k) dk.
\end{align}
In particular,
\begin{align}\label{hu}
(\gr, \s \phi(h) \gr)=
\alpha \int_\BR \| (H-E+\omega(k))^{-1/2}\s\gr\|^2   |\hat h(k)|^2 dk.
\end{align}
\ec
\proof
By Theorem \ref{expectationoffield2}
we have
\begin{align*}
(\gr, \xi(\s) \phi(f) \gr)&=\EEEE \lkkk \xi(X_0) \KI  \rkkk
=\frac{\alpha}{2}\int_{-\infty}^\infty dr (e^{-|r|\omega}\hat h, \hat f)
\EEEE \lkkk \xi(X_0) X_{\eps r}\rkkk.
\end{align*}
By Corollary \ref{expectation} we also see that this furthermore is
\begin{align*}
&=\frac{\alpha}{2}\int_{-\infty}^\infty dr \int dk
(\xi(\s)\gr, e^{-|r|(H-E+\omega(k))}\s\gr) \ov{\hat h(k)}\hat f(k)\\
&=
\alpha
\int_\BR (\xi(\s)\gr,
 (H-E+\omega(k))^{-1}\s\gr) \ov{\hat h(k)}\hat f(k)dk.
\end{align*}
Then the corollary follows.
\qed
A standard inequality, see e.g. \cite[Proposition 5.1]{lhb11}, says that
$$
(\Phi, (\s\phi(f))^2\Phi)\leq \frac{\|f\|^2}{2}(\Phi, (N+\one)\Phi).
$$
From Corollary \ref{yume3}  we obtain the following inequality.
\bc{ine}
Let $\hat h/\omega\in\LR$. Then
\eq{ine1}
(\gr, N\gr)\leq \frac{\alpha}{2}(\gr, \s\phi(\omega({\rm D})^{-1}h)\gr)\leq
\frac{\alpha^2}{2}\|\hat h/\omega\|^2,\quad {\rm D}=-i\nabla.
\en
\ec
\proof
By
\kak{m} we have
$$
(\gr, N \gr)=\frac{\alpha^2}{2}\int_\BR|\hat h(k)|^2
\int_{-\infty}^0dt\int_0^\infty ds e^{-|t-s|\omega(k)}\EEEE \lkkk X_{\eps t} X_{\eps s}\rkkk.
$$
Since  $\EEEE [X_ {\eps s}X_{\eps  t}]=(\s\gr, e^{-|t-s|(\PF-E)}\s\gr)$, by
Corollary \ref{finitedistribution2} we see that
\eq{pt}
(\gr, N\gr) = \frac{\alpha^2}{2} \int_{\BR} {|\hat h(k)|^2}
\| (\PF-E+\omega(k))^{-1}\s\gr\|^2 dk.
\en
The first inequality is derived from  \kak{pt} and \kak{hu}. The second inequality follows through
\kak{hu}.
\qed

Since $\gr\in D(N)$, we have  that $f\mapsto (\gr, \s\phi(f) \gr)$ is linear and the bound
$|(\gr, \s\phi(f) \gr)|\leq C\|f\|$ with a constant $C$ follows. By the Riesz representation theorem
there exists $G\in\LR$ such that $(\gr, \s\phi(f)\gr)=(\hat G,\hat f)_\LR$.
\bc{yume4}
If $\hat h/\omega\in\LR$, then
\eq{gk}
\hat G(k)=
\frac{\alpha}{2}
(\s\gr, (H-E+\omega(k))^{-1}\s\gr){\hat h}(k).
\en
\ec
\proof
This is obtained directly from Corollary  \ref{yume3}.
\qed

\subsection{Comparison with the Nelson model}
\label{L2}
From \kak{gk} we see that, formally, $G(\cdot)=(\gr, \s\phi(\cdot )\gr)$. Recall that the Nelson
model is defined by a linear coupling between a particle described by the Schr\"odinger operator
$H_{\rm p}= -\Delta+V$ and a boson field described by $\hf$. The coupling term is given by
\eq{nelson2}
\phi_{\rm b}(x)=\frac{1}{\sqrt 2}\int \left(\add(k) e^{-ikx}{\hat h(-k)}+a(k) e^{-ikx}\hat h(k)\right) dk
\en
so that the Nelson Hamiltonian is defined by
\eq{nelson1}
H_{\rm N} =H_{\rm p} +\hf+\phi_{\rm b}(x).
\en
For this model a similar kernel to \kak{gk} is obtained, see \cite[eq. (6.5.60)]{lhb11}. Using this kernel
we can show \cite[Sect. 6.5]{lhb11} that
\eq{neline}
a\|\hat h/\omega\|^2-b\leq (\gr^{\rm N}, N \gr^{\rm N})\leq c\|\hat h/\omega\|^2,
\en
where  $\gr^{\rm N}$ is the ground state of $H_{\rm N}$, and $a,c\geq0$ and $b\in\RR$ are suitable constants.
Thus $(\gr^{\rm N}, N \gr^{\rm N})\to\infty$ as $\|\hat h/\omega\| \to \infty$, and it follows that
$H_{\rm N}$ has no ground state whenever $\hat h/\omega\not\in \LR$. The key point is that $(\gr^{\rm N},
e^{i(\cdot,x)}\gr^{\rm N})\not =0$; for some discussions see \cite{ahh99, ghps12}. While in \kak{ine1} we show the upper
bound $\d (\gr, N\gr)\leq  \frac{\alpha^2}{2} \|\hat h/\omega\|^2$, we have no lower bound like in \kak{neline}
due to the fact that $(\gr, \s\gr)=0$. Although there is no mathematical proof known, the physics literature
supports the conjecture that whenever a ground state exists, the boson number expectation 
$(\gr^{\rm N}, N\gr^{\rm N})$ is finite. Thus $\hat h/\omega \not \in L^2(\BR)$ would then mean absence  of a 
ground state. For the spin boson model it can be conjectured for similar physical reasons that the same mechanism 
applies and $\hat h/\omega\not \in L^2(\BR)$ implies absence of a ground state. Then some interesting open questions 
are if it is possible to prove $(\gr, N\gr) \to \infty$ as $\|\hat h/\omega\|\to\infty$ and, secondly, if a ground 
state of $H$ exists when $\hat h/\omega \not\in\LR$. These questions will be considered elsewhere.

\bigskip\bigskip
\noindent
{\bf Acknowledgments:} This work was financially supported by Grant-in-Aid for Science Research (B) 23340032
from JSPS. JL thanks IHES, Bures-sur-Yvette, for a visiting fellowship.

\end{document}